# Ferromagnetic resonance modes in the exchange dominated limit in cylinders of finite length

Jinho Lim[1], Anupam Garg[1] and J. B. Ketterson[1,2]

1. Department of Physics and Astronomy, Northwestern University, Evanston, IL, 60208.
2. Department of Electrical and Computer Engineering Northwestern University, Evanston, IL, 60208.

## Abstract

We analyze the magnetic mode structure of axially-magnetized, finite-length, nanoscopic cylinders in a regime where the exchange interaction dominates, along with simulations of the mode frequencies of the ferrimagnet yttrium iron garnet. For the bulk modes we find that the frequencies can be represented by an expression given by Herring and Kittel by using wavevector components obtained by fitting the mode patterns emerging from these simulations. In addition to the axial, radial, and azimuthal modes that are present in an infinite cylinder, we find localized "cap modes" that are "trapped" at the top and bottom cylinder faces by the inhomogeneous dipole field emerging from the ends. Semi-quantitative explanations are given for some of the modes in terms of a one-dimensional Schrodinger equation which is valid in the exchange dominant case. The assignment of the azimuthal mode number is carefully discussed and the frequency splitting of a few pairs of nearly degenerate modes is determined through the beat pattern emerging from them.

## 1. Introduction

Measurements of the resonant microwave response of the spins of unpaired electrons in ferromagnetic materials have a long history, beginning with its initial observation by Griffiths[1]. The vast majority of the subsequent experimental studies have involved "macroscopic" samples where the spins excited by an external microwave field precess in the combined field arising from an externally applied constant field together with the field produced by the sample's own magnetization (the so-called demagnetization field) under conditions where both are nearly uniform. Under such conditions the strength of the exchange interaction, though itself responsible for the materials magnetization, does not affect the resonance frequency.

In the decade after Griffiths' work, it was found that attempts to drive the resonance to larger amplitudes did not work, and that it saturated at values of the microwave power far below that expected from the observed resonance linewidth. The explanation is that small inhomogeneities in the magnetization of the sample are amplified by the inhomogeneities generated in turn in the dipolar or demagnetizing field, and leads to what are now called the Suhl instabilities[2,3]. In addition to studies of the uniformly precessing mode, spatially-nonuniform modes have also been intensively studied, both experimentally and theoretically, as we shall describe below.

In this paper, we report on an exhaustive numerical study, using the OOMMF micromagnetic simulation code[4], of the resonance modes of Yttrium Iron Garnet (YIG) cylinders, primarily of diameter d = 75nm and height h = 300 nm, although some aspects have been studied for other values of h (7.5–1200 nm). Our methodology shares many features with the work of McMichael and Stiles[5] on two-dimensional elliptical disks and three dimensional thin cylindrical discs. Our three-dimensional geometry displays a much richer mode structure, however, requiring a more detailed theoretical framework. In addition, we have also developed techniques to resolve modes that are nearly degenerate in frequency.

The motivation for our work is as follows. Firstly, in recent years many authors have examined whether the magnetization of small particles (in the size range of order $10^2$ nm) can be reversed by applying microwave fields (so-called microwave assisted switching) with a view to applications in magnetic storage[6,7,8,9]. The reasoning is that at small sizes, the exchange interaction favors parallel alignment of the spins and suppresses the dipolar instabilities. In previous work, we have used OOMMF to simulate YIG cylinders with (d, h) = (25, 50)nm, (25,100)nm and (75,150)nm, where we have attempted to reverse the magnetic moment of the cylinder by applying pi-pulses[10,11], as in NMR[12]. We found that by chirping the frequency of the pulse, we could achieve reversal in the 25 nm diameter samples, but not in the larger ones. If pi-pulses or other reversal protocols are to be successfully implemented in more realistic larger samples, then an understanding of the normal modes in restricted geometries is a necessary prior step. The problem is also of intrinsic interest, and we have found novel features not seen in ellipsoidal geometries (including degenerate forms thereof) that have been the subject of almost all other work to date.

Secondly, with recent advances in sub-micron patterning techniques, the study of arrays of objects (so as to have large signals) for which the largest dimension is few hundred nanometers or less is attracting increasing attention. With advanced techniques it is even possible to probe the magnetic properties of individual sub-micron particles[13,14,15]. Measurements on such samples can even be performed in the absence of an external field, i.e., solely in the presence of the internal demagnetization field (for shapes where such a field exists), provided the sample is small enough to be in a single domain state[16]. Modes with an odd number of maxima and minima can be excited directly with a uniform microwave field; coupling to modes with higher wave numbers will be more challenging[17].

Here we will primarily be concerned with *size-quantization effects* arising from finite sample dimensions. In particular we will examine the mode spectrum in samples having a cylindrical shape with radius a (and corresponding diameter d = 2a) and height h, both analytically and numerically. Due to the ease of preparation of some materials as wires, such samples are widely studied experimentally, e.g., in permalloy Py[18] and in Ni[19,20]. Cylinders of finite length with h/d ratios of order unity and larger can be readily patterned using optical and e-beam lithography by creating hole arrays in a resist followed by deposition and liftoff[21].

## 1.1. Theoretical background

Free spins in a magnetic field H precess at the Larmor frequency, $\omega = \gamma H$, where $\gamma = g|e|/2mc$ with g, e, and m being the electron g-factor, charge and mass. As noted, in materials having an internal magnetization additional fields are present which can alter the precession frequency. To describe this and related effects, Landau and Lifshitz[22] (LL) introduced the following equation of motion

$$\frac{d\mathbf{M}}{dt} = -\gamma \mathbf{M} \times \mathbf{H} - \frac{\alpha\gamma}{M_0} \mathbf{M} \times (\mathbf{M} \times \mathbf{H}) ; \qquad (1.1)$$

here $\mathbf{H}$ is the total field at a given position within the sample arising from the external field as well as that produced by the magnetization itself and an effective field arising from quantum mechanical exchange; it can also include crystalline anisotropy, but this is suppressed in what follows. The second term on the right-hand side of Eq. (1) is incorporated to phenomenologically account for damping, which will largely be neglected in what follows. In addition to satisfying Eq. (1) $\mathbf{M}$ and $\mathbf{H}$ must satisfy appropriate boundary conditions at the surface of the body.

For ellipsoidal samples (including degenerate forms thereof), and in the presence of a homogeneous external field $\mathbf{H}$, the magnetization $\mathbf{M}$ is nominally homogeneous as is the resulting demagnetization field; one can then observe sharp absorption lines in ferrommagnetic resonance (FMR) experiments (in the absence of strong damping), all spins then seeing the same local field. The resonance frequency of this uniformly precessing mode in a spheroidal sample (where two of the principal axes of the ellipsoid are identical) with the external field $H_0$ along the axis of rotation is given by what is commonly called the Kittel formula,[23]

$$\omega = \gamma \left( H_0 + 4\pi (N_\perp - N_\parallel) M_0 \right), \tag{1.2}$$

where $N_\perp$ and $N_\parallel$ are coefficients accounting for the effect of demagnetization perpendicular and parallel to the rotation axes (with $2N_\perp + N_\parallel = 1$), and $M_0$ is the internal magnetization, taken as a constant; note $\gamma$ may differ from the free-space value due to atomic and solid-state effects.

In addition to the uniformly precessing mode there exist non-uniform modes[24] which we can characterize by some effective wavelength, $\lambda$. At shorter (nanometer) scale wavelengths, the exchange interaction dominates, and the associated modes are termed exchange modes, first introduced by Bloch[25]. The importance of modes with longer wavelengths (in suitably large samples), was emphasized by Clogston, Suhl, Walker and Anderson[26,27]. They arise from a solution of Eq. (1) together with $\nabla \cdot \mathbf{B} = 0$ and the Maxwell boundary conditions; they are commonly referred to as *magnetistatic modes*. Modes in the region where both exchange and magnetostic effects compete are called dipole/exchange modes.

For the case of a sphere some of the low-lying magnetostaic modes were first examined by Mercerau and Feynman[28]. They were later studied in much greater detail for spheroidal samples by Walker[29].

The limiting case of a finite thickness, infinite-area, slab ($N_\perp = 0, N_z = 4\pi$) with both the wave vector $\mathbf{k}$ ($|k| = 2\pi / \lambda$) and the external magnetic field $\mathbf{H}_0$ lying in plane was treated by Damon and Eshbach[30]; the case with $\mathbf{H}_0$ perpendicular to the film was examined by Damon and Van De Vaart[31]. For an in-plane field Damon and Eshbach identified two classes of magnetostatic modes: a surface wave and a family of bulk waves. The first of these, now designated as the DE mode, decays exponentially within the interior. The DE mode is most commonly studied for wave vectors $\mathbf{k} \parallel (\mathbf{H}_0 \times \hat{\mathbf{z}})$ where $\hat{\mathbf{z}}$ is the plane normal; it has a positive group velocity but also has the unusual property that it propagates only on *one side* of the slab for a given field direction, switching to the opposite side on reversing the field. The second class is a family of bulk modes which are quantized in the film thickness direction; they are typically studied with $\mathbf{k} \parallel \mathbf{H}_0$. They propagate in both directions and the lowest lying mode has the unusual property that its group velocity is negative for small k and therefore it is called a *backward volume mode*. The frequencies of all the volume modes asymptotically approach the free spin precession frequency, $\omega = \gamma H_0$, as $k \to \infty$ (in the absence of exchange); they approach the in-plane Kittel frequency, $\gamma \sqrt{(H_0(H_0 + 4\pi M_0)}$ , as $k \to 0$.

For $H_0$ perpendicular to the film we again have modes quantized along the film thickness that now propagate isotropically in plane. The lowest has a positive group velocity for small in-plane wave vector k; it is therefore referred to as a *forward volume mode*. All these modes approach the perpendicular Kittel frequency $\gamma(H_0 - 4\pi M)$ as $k \to 0$ and $\gamma \sqrt{(H_0(H_0 - 4\pi M_0)}$ as $k \to \infty$. Exchange effects have been considered by De Wames and Wolfram[32] and more completely by Arias[33].

For the case of an infinitely long cylinder ( $N_\perp = 2\pi; N_z = 0$ ) with $H_0$ parallel to the rotational axis, which is relevant to the work presented here, the mode structure was first studied by Joseph and Schlomann[34]. Here we encounter families of purely azimuthal modes and analogous to the backward volume modes we have radially quantized modes propagating up and down the cylinder axis, which also approach $\omega = \gamma H_0$ at large k (in the absence of exchange). Recently this problem was reexamined by Arias and Mills[35] who also considered the effects of exchange via perturbation theory.

At shorter wavelengths the effects of exchange contribute. In this regime the frequency of a mode with wave vector $k = 2\pi/\lambda$ for a spheroidal sample with the external field $H_0$ aligned along the rotational axis can be described by the Herring-Kittel formula,[36] which we discuss in Appendix A

$$\omega = \gamma \sqrt{(H_0 - 4\pi N_\parallel M_0 + D_{ex}k^2)(H_0 - 4\pi N_\parallel M_0 + D_{ex}k^2 + 4\pi M_0 \sin^2\theta)} \; ; \qquad (1.3)$$

here $D_{ex}$ is a parameter measuring the strength of the exchange (see below), $k^2 = k_z^2 + k_\perp^2$, $k_z$ and $k_\perp$ are the components of the wavevector parallel and perpendicular to the spheroid axis, and $\theta = \tan^{-1}(k_\perp/k_z)$ is the angle between the spin wave propagation direction and the spheroid axes. Note that at $k = 0$ the factor involving $N_\perp$ that appears in Eq. (1.2) *is absent* from Eq. (1.3), since it is assumed that the transverse demagnetization field is "screened out" at short wavelengths. Indeed $\omega$ is ill defined at precisely $k = 0$ since $\theta$ is ambiguous. This shortcoming also signals the importance of the magnetostatic modes at intermediate k values; i.e., as the sample size is reduced there is a crossover between dipole dominated and exchange dominated modes. Modes with k values straddling these regimes are the dipole-exchange modes mentioned above. At short wavelengths, which will be the case in sufficiently small samples, Eq. (1.3) should provide a representation of the mode structure in rotationally symmetric samples, provided quantized values of $k_\perp$ and $k_z$ satisfying the boundary conditions are available; we will utilize Eq. (1.3) to represent some of our finite-size sample simulations in what follows.

More generally and in the absence of exchange effects, magnetostatic effects would dominate the mode frequencies, which for a spheroid would lie in the range

$$\gamma(H_0 - 4\pi N_\parallel M_0 + 2\pi M_0) \geq \omega \geq \gamma(H_0 - 4\pi N_\parallel M_0) ; \qquad (1.4)$$

note the number of modes in this interval is bounded only by the number of spins; i.e., the mode density is very high, making the resolution of the individual modes extremely difficult at shorter wavelengths (where they pile up). When exchange is present the mode frequencies are spread over a much wider interval.

In an inhomogeneous external field, or for samples with an arbitrary shape, one might initially expect to observe a line-broadening (as happens in most nuclear magnetic resonance experiments). However, in the presence of exchange this is not the case and well-defined modes emerge as will be discussed further below. Exchange modes are not restricted to magnetic materials; they have been observed in experiments on Fermi liquids at low temperatures, examples being certain metals[37] and the normal state of liquid $^3$He [38]; the theory for the latter was first given by Silin[39] for homogeneous fields and later extended by Leggett[40] to describe the inhomogeneous case.

1.2. Plan of the paper

We develop the theoretical framework for our problem in Sec. 2, beginning with a discussion of cylindrical symmetry and the resulting angular momentum quantum number (or azimuthal mode number) in Sec. 2.1. In the magnetostatic limit it is convenient to take this as the total angular momentum, m, as done by Walker, and by Joseph and Schlomann[34]. In the exchange dominated limit, it becomes more important to understand the separation of the angular momentum into its orbital and spin parts. The major component of a mode has spin $m_s = 1$ and orbital angular momentum $m_l = p$, and there is a small admixture of $m_s = -1$ and $m_l = p + 2$. Accordingly, we find it better to label the modes by the orbital angular momentum p of the major component. This is especially so when examining the computer-generated mode patterns since the orbital behavior of any component of $\mathbf{M}$ is immediately apparent.

That the two components of a mode are so unequal goes hand in hand with the fact that modes with $m_l$ = p and −p are nearly degenerate, as we discuss in Sec. 2.2. A clear understanding of this issue is important, as this near degeneracy can lead to some confusion when looking at the mode patterns. In one case, we have resolved this degeneracy (see Sec. 3.4) by exciting and examining the beat pattern between the ±p modes.

In Sec. 2.3, we show that exchange dominated modes in long cylinders are approximately described via a Schrodinger-like equation for a particle in a cylindrical box with a modified boundary condition, such that axial and radial dependence of the mode function factorizes, and the resulting quantization gives rise to axial and radial mode numbers. In an infinite cylinder this separation is exact, which is exploited to good effect in the analyses of Joseph and Schlomann[34] and of Arias and Mills[35]. In a finite cylinder, the separation is approximate since the demagnetizing field is non-uniform, and flares away from the axis near the perimeter of caps at z = 0 and z = h. We give a semi-quantitative argument in Sec. 2.4 that the Schrodinger equation possesses bound state solutions near these caps, corresponding to "cap modes" which we see very clearly in our simulations. For any p, there are two such modes (one for each cap), whose frequencies lie below those of the bulk modes with the same p. This means that the uniform FMR or Kittel mode, which is the lowest bulk mode with p = 0, is not the lowest frequency mode of the body. For this case, we present numerical results for the solution of the Schrodinger equation in Sec. 2.5, and find good agreement with the simulations.

The cap modes are a novel and unexpected feature of our study, as they do not exist in an infinite cylinder or a finite sized ellipsoid of revolution. Similar "end modes" were found by McMichael and Stiles[5], who did not however investigate their origin. We expect that such localized modes will exist near the surfaces of other sample shapes as well whenever the demagnetizing field departs significantly from uniformity.

Our simulational approach is described in Sec. 3. It is based on the OOMMF code developed at the National Institute of Standards and Technology. After finding the static equilibrium magnetization, $\mathbf{M}_{eq}(\mathbf{r})$ (Sec. 3.1), we can excite the system by applying pulses that are localized in either space, time or both[41]. The spatial center and the width and frequency bandwidth are varied depending on which mode(s) we wish to excite. The resulting time development of $\mathbf{M}(\mathbf{r},t)$ is Fourier transformed, and the point-wise power spectrum is added over all the cells. The resulting sum displays peaks at many mode frequencies, and by honing in on individual peaks, we can identify the magnetization patterns for each mode as explained in Sec. 3.3.

Once a particular mode pattern is obtained in a simulation it can be used as is or altered in some way, say by combining it with some other mode, to study the subsequent development in time. This is a potentially promising way to study mode-mode coupling or large amplitude responses which we hope to pursue in the future. As an application of this idea, and as noted above, ±p modes are sometimes nearly degenerate, as are the even and odd super-positions of the cap modes. In Sec. 3.4 we show that by starting the simulation in a suitable real-space pattern we can find a beat pattern in the time development of the magnetization from which we can obtain the frequency splitting of the modes. We have performed this exercise for only a few cases as it is computationally intensive, and the physical principles are the same for the other cases.

In Sec. 4 we tabulate the frequencies of all the modes we have found (approximately 90) and discuss the assignment of mode numbers further. The assignment of the longitudinal quantum number $n_z$ on the basis of the one-dimensional Schrodinger equation is particularly tricky as the existence of the cap modes forces nodes in the bulk mode functions near the caps, and prevents accurate fitting of the lowest few bulk modes to a sinusoidal form $\sin(k_z z)$ with $k_z$ strictly equal to $\pi/h$ times an integer. Nevertheless, an unambiguous labeling of the modes is possible.

In Sec. 5 we show that the mode frequencies that we obtain agree surprisingly well with the Herring-Kittel expression (1.3) provided we identify $k_\perp$ and $k_z$ in this formula correctly. We give reasons why this agreement might be so good, explain how the wavevector components are found, and how this allows us to organize the normal mode spectrum into families of modes labeled by p.

Spatial patterns for a variety of modes are given in Sec's. 6 and 7 (in Figs. 6.1–6.8 and Fig. 7.2). These patterns are the centerpiece of our paper, and show beautiful regularity and symmetry. In Sec. 6 we consider only the d = 75 nm, h = 300 nm sample, while in Sec. 7 we consider the lowest three p = 0 modes as a function of h. We find that at small h/d (disc-like sample), the symmetric cap mode (which has the lowest frequency of all three) is in fact the mode that one would regard as the uniform FMR or Kittel mode, and its frequency is well fit by the Kittel formula with an appropriate choice of demagnetization coefficients. For large h/d however, it is the lowest bulk mode (whose frequency lies above the two cap modes) that should be identified with the Kittel mode. For intermediate values of h/d $\simeq$ 6–8, the Kittel formula does not actually describe any of the modes. To our knowledge this point has not been appreciated before. Once again it illustrates the richness of the normal mode spectrum in non-ellipsoidal samples.

Finally, Sec. 8 summarizes our conclusions. Here we take the opportunity to emphasize the importance that simulations of the small amplitude mode structure in nano-structures have for present and possible future applications, some of which are currently speculative in character.

## 2. Modes in the exchange dominated limit

In the presence of an isotropic exchange interaction, and neglecting the effects of damping, Eq. (1.1) takes the form[42]

$$
\begin{aligned}
\frac{d\mathbf{M}}{dt} &= -\gamma \, \mathbf{M} \times \left( \mathbf{H} - \frac{2A_{ex}}{M_0^2} \nabla^2 \mathbf{M} \right) \\
&= -\gamma \, \mathbf{M} \times \left( \mathbf{H} - \frac{D_{ex}}{M_0} \nabla^2 \mathbf{M} \right),
\end{aligned}
\tag{2.1}
$$

where $A_{ex}$ is a parameter fixing the strength of the exchange interaction and $D_{ex} \equiv 2A_{ex}/M_0$ Here $\mathbf{H}$ is the applied magnetic field, $H_0\hat{\mathbf{z}}$, plus the dipolar or demagnetizing field generated by $\mathbf{M}$. In a cylinder of finite height, the dipolar field is not uniform, especially near the caps, and so the static equilibrium field, $\mathbf{M}_{eq}(\mathbf{r})$ is not everywhere parallel to $\hat{\mathbf{z}}$. A linearized normal mode analysis should therefore consider deviations $\delta\mathbf{M}(\mathbf{r},t) \perp \mathbf{M}_{eq}$, which do not lie in the x-y plane. If exchange is strong, however, the non-uniformity in $\mathbf{M}_{eq}$ is very small (this is true for all the simulations we have performed), and we may then take $\delta\mathbf{M}_z = 0$. This assumption makes it much easier to discuss the physics, and relaxing it only obscures the key ideas without adding substance. We stress that it is not essential to our argument, especially with respect to the symmetries and the azimuthal quantum number. With this assumption, we may write

$$
\mathbf{M}(\mathbf{r},t) = M_0\left((1-m^2)^{1/2}\hat{\mathbf{z}} + \mathbf{m}(\mathbf{r},t)\right),
\tag{2.2}
$$

where $\mathbf{m}$ has only x and y components and is dimensionless since we have scaled out $M_0$.

For small deviations, $|\mathbf{m}| \ll 1$, the linearized LL equation can be cast as

$$\frac{d\mathbf{m}(\mathbf{r},t)}{dt} = \gamma\,\hat{\mathbf{z}} \times \left( H_z(\mathbf{r},z)\mathbf{m}(\mathbf{r},t) - M_0\mathbf{h}_d(\mathbf{r},t) - D_{ex}\nabla^2\mathbf{m}(\mathbf{r},t) \right). \qquad (2.3)$$

Here $H_z(\mathbf{r},z)$ consists of the applied field, $H_0\hat{\mathbf{z}}$, together with the position dependent longitudinal demagnetization field arising from the static magnetization, and $\mathbf{h}_d$ is the (small) demagnetization field induced by $\mathbf{m}$. (We use cylindrical coordinates $\mathbf{r} = (r,\varphi,z)$ here and below.)

### 2.1 Assignment of the angular momentum quantum number

Equation (2.3) defines an eigenvalue problem with cylindrical symmetry, so there must exist solutions with definite azimuthal mode number. In the zero-exchange or magnetostatic limit, the analysis is best done in terms of a *scalar* magnetic potential $\psi$, which varies as $e^{im\varphi}$ in the eigenmodes; the integer m (which we must be careful to distinguish from the scalar value of $\mathbf{m}$) is then naturally interpreted as the angular momentum quantum number. In the strong exchange limit, the problem is better formulated in terms of $\mathbf{m}$ directly, which as a *vector* field transforms differently under rotations than a scalar field[43] (such as $\psi(\mathbf{r})$ in the Schrodinger equation).

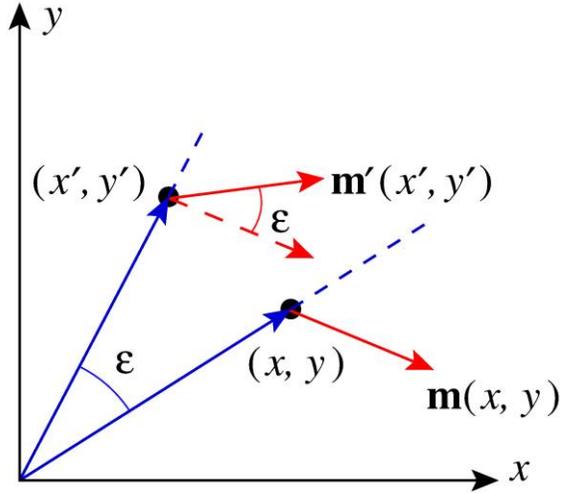

**Fig. 2. 1. Transformation of a vector field.**

Let us examine the effect of a rotation on the vector $\mathbf{m}$ at a point $(x,y,z)$ by an angle $\varepsilon$ about the z axis to a vector $\mathbf{m}'$ at the point $(x',y',z)$. We need only carry this analysis out to leading order in $\varepsilon$. The components of the rotated vector $\mathbf{m}'$ are then (see Fig. 2.1.)

$$m_x'(x',y',z) = m_x(x,y,z) - \varepsilon m_y(x,y,z), \qquad (2.4a)$$

$$m_y'(x',y',z) = \varepsilon m_x(x,y,z) + m_y(x,y,z). \qquad (2.4b)$$

The coordinates $(x',y',z)$ themselves are related to $(x,y,z)$ as

$$x' = x\cos\varepsilon + y\sin\varepsilon = x - \varepsilon y, \qquad (2.5a)$$

$$y' = y\cos\varepsilon + x\sin\varepsilon = y + \varepsilon x. \qquad (2.5b)$$

If we expand the left side of (2.4a,b) to first order in $\varepsilon$, note that to zeroth order

$\mathbf{m}'(x',y',z) = \mathbf{m}(x,y,z)$, and recall the definition of the (dimensionless) orbital angular momentum operator $l_z$ in quantum mechanics as

$$l_z = -i\left(x\frac{\partial}{\partial y} - y\frac{\partial}{\partial x}\right) = -i\frac{\partial}{\partial \varphi}, \tag{2.6}$$

we can write $\mathbf{m}'$ in terms of $\mathbf{m}$ as

$$m'_x(x,y,z) = (1 - i\varepsilon l_z)m_x(x,y,z) - i\varepsilon[-i\,m_y(x,y,z)], \tag{2.7a}$$

$$m'_y(x,y,z) = (1 - i\varepsilon l_z)m_y(x,y,z) - i\varepsilon[i\,m_x(x,y,z)]. \tag{2.7b}$$

Equation (2.7) can be rewritten in the form

$$\begin{pmatrix} m'_x \\ m'_y \end{pmatrix} = \begin{pmatrix} 1 - i\varepsilon l_z & 0 \\ 0 & 1 - i\varepsilon l_z \end{pmatrix}\begin{pmatrix} m_x \\ m_y \end{pmatrix} - i\varepsilon\begin{pmatrix} 0 & -i \\ i & 0 \end{pmatrix}\begin{pmatrix} m_x \\ m_y \end{pmatrix}. \tag{2.8}$$

For a *scalar* field, $\psi(\mathbf{r})$, we would simply have $\psi'(\mathbf{r}) = (1 - i\varepsilon l_z)\psi(\mathbf{r})$, but the presence of the last term in (2.8) *mixes* the two components of the *vector* field $\mathbf{m}$. This can be interpreted as arising from an "internal" or "spin" angular momentum of $m_s = \pm 1$ that is added to or subtracted from the orbital angular momentum $m_l$ associated with our vector field $\mathbf{m}$ (a tensor of rank 1). A similar separation exists in the description of light fields[44].

Consider the case of a vector field of the form

$$m_x(\mathbf{r}) = a(r,z)e^{ip\varphi}, \qquad m_y(\mathbf{r}) = b(r,z)e^{ip\varphi} \qquad \text{(a, b arbitrary)}. \tag{2.9}$$

This field has orbital angular momentum $m_l \equiv p$, but has no definite spin angular momentum. For it to have a definite spin $a$ and $b$ must be proportional according to

$$\begin{pmatrix} m_x \\ m_y \end{pmatrix} \propto \begin{pmatrix} 1 \\ i \end{pmatrix} \Leftrightarrow m_s = 1 \qquad \text{or} \qquad \begin{pmatrix} m_x \\ m_y \end{pmatrix} \propto \begin{pmatrix} 1 \\ -i \end{pmatrix} \Leftrightarrow m_s = -1. \tag{2.10a, b}$$

Writing $m_\pm = m_x \pm im_y$ it then follows that

$$m_s = 1 \text{ yields } m_+ = 0, m_- \propto e^{ip\varphi} \text{ and } m_{tot} = p + 1 \tag{2.11a}$$

and

$$m_s = -1 \text{ yields } m_+ \propto e^{ip\varphi}, m_- = 0 \text{ and } m_{tot} = p - 1 \tag{2.11b}$$

where we wrote $m_{tot} = m_l + m_s$. It follows that an eigen mode with total angular momentum $p + 1$ must be of the form

$$\begin{pmatrix} m_x(\mathbf{r},t) \\ m_y(\mathbf{r},t) \end{pmatrix} = F_-(r,z)\begin{pmatrix} 1 \\ i \end{pmatrix}e^{i(p\varphi - \omega t)} + F_+(r,z)\begin{pmatrix} 1 \\ -i \end{pmatrix}e^{i((p+2)\varphi - \omega t)} \tag{2.12}$$

where we have adopted an $e^{-i\omega t}$ time dependence following standard practice. The physical solution is obtained by taking the real part of this complex-valued solution. We shall see below that for positive frequency solutions, $F_- \gg F_+$ in the strong exchange limit, and it is often convenient to neglect $F_+$ entirely. It is then more useful to label the modes by p, the orbital angular momentum of the dominant component, $F_-$. This is especially so when looking at mode patterns generated by OOMMF, since we can read off p by seeing how many times $\mathbf{m}$ turns as we go around a circle in the x-y plane. For example, in a $p = 0$ mode (see Figs. 6.1a and 6.1b) $\mathbf{m}$ appears uniform, while in $p = -1$ (Fig. 6.3) and $p = 1$ (Fig. 6.4) modes $\mathbf{m}$ winds by $2\pi$ and $-2\pi$, respectively as we go anti-clockwise around a circle.

In the magnetostatic limit by contrast, $F_+ / F_- \sim O(1)$, and the m labeling is better. Thus, for the sphere, while we would describe the uniform or Kittel mode as having $p = 0$, Walker[29] assigns m = 1 to it (see his Fig. 3 where the mode is labeled (110)).

## 2.2. Near degeneracy of p and − p modes

When exchange dominates over dipole-dipole interactions, we may as a first approximation neglect $\mathbf{h}_d$ in Eq. (2.3). In component form the equation then reads

$$\frac{d}{dt}\begin{pmatrix} m_x(\mathbf{r},t) \\ m_y(\mathbf{r},t) \end{pmatrix} = \gamma\left(H(r,z) - D_{ex}\nabla^2\right)\begin{pmatrix} -m_y(\mathbf{r},t) \\ m_x(\mathbf{r},t) \end{pmatrix} \tag{2.13}$$

or

$$\frac{d}{dt}m_\pm(\mathbf{r},t) = \pm\gamma\left(H(r,z) - D_{ex}\nabla^2\right)m_\pm(\mathbf{r},t) \tag{2.14}$$

with

$$m_\pm(\mathbf{r},t) = m_x(\mathbf{r},t) \pm im_y(\mathbf{r},t). \tag{2.15}$$

We seek solutions of the form

$$m_\pm(\mathbf{r},t) = m_\pm(\mathbf{r})e^{-i\omega t} \tag{2.16}$$

and demand that $\omega > 0$, with the understanding that the physical solution will be given by the real part. These solutions then obey

$$\gamma\left(H_z(r,z) - D_{ex}\nabla^2\right)m_{\pm}(\mathbf{r}) = \mp\omega m_{\pm}(\mathbf{r}). \tag{2.17}$$

This equation is like a one-particle Schrodinger equation, and since $\gamma > 0$ in our convention, the operator on the left is a positive operator which cannot have negative eigenvalues. Since we also demand that $\omega > 0$, we must choose $m_+ = 0$. Finally, since our finite cylinder retains full azimuthal symmetry, the solution for $m_-$ takes the form

$$m_-(r,\varphi,z) = F_-(r,z)e^{ip\varphi}, \tag{2.18}$$

where $F_-(r,z)$ can be chosen to be real, and $\omega$ has the same (positive) value for either sign of p. In terms of the general form (2.12), this solution corresponds to putting $F_+ = 0$, and a physical solution

$$\begin{pmatrix} m_x(\mathbf{r},t) \\ m_y(\mathbf{r},t) \end{pmatrix} = F_-(r,z)\begin{pmatrix} 1 \\ i \end{pmatrix}e^{i(p\varphi - \omega t)} + \text{c.c.} = 2F_-(r,z)\begin{pmatrix} \cos(p\varphi - \omega t) \\ \sin(p\varphi - \omega t) \end{pmatrix}. \tag{2.19}$$

If we now include the dipolar field $\mathbf{h}_d$ as a perturbation, we can expect that $F_+$ will become nonzero, with $F_+ / F_- \sim 4\pi M_0 / D_{ex}k^2$, where $2\pi / k$ is the typical length scale on which the solution varies.

The source of the degeneracy with respect to $\pm p$ is that Eq. (2.13) is invariant under reflection in the yz plane provided we do not also reflect the vector $\mathbf{m}$. Hence the operation

$$m_x(x,y,z) \to m_x(-x,y,z) \tag{2.20a}$$

and

$$m_y(x,y,z) \to m_y(-x,y,z) \tag{2.20b}$$

also produces a solution. This operation is equivalent to $\varphi \to -\varphi$, or alternatively to $p \to -p$. Inclusion of the dipole-dipole interaction destroys this invariance: the field $\mathbf{h}_d$ produced by the operation is not the same field as before.

Strictly speaking therefore, modes differing only in the sign of p are not degenerate, although the non-degeneracy may be small. Indeed, as explained below, we have spent significant effort to numerically resolve the splitting and have not always succeeded. The physical origin of this non-degeneracy is just that the applied external field breaks time reversal and parity symmetries. In the magnetostatic limit, this point emerges directly from the solution in terms of the scalar potential. Joseph and Schlomann[45] find that $\omega_{-|m|} > \omega_{|m|}$ for the volume modes (where we have here used m instead of p to label the modes), and that only m > 0 solutions exist for the surface modes. This is an extreme form of the nondegeneracy, and is the cylindrical analog of Damon and Eshbach's discovery of one-sided surface modes in the slab geometry. Joseph and Schlomann[34] also find that the ±p splitting becomes smaller with increasing $k_z$ or increasing radial mode number (see their Fig. 5). The same behavior is found for the general spheroid by Walker.[46] Arias and Mills[35] on the other hand, appear to us to be finding that modes with the opposite sign of the angular momentum are degenerate; we are unable to pinpoint why.

The near degeneracy of ±p modes also underlies whether one sees azimuthal standing or running wave patterns in the OOMMF simulations. We discuss this issue in Sec. 3.3 below.

## 2.3. The long cylinder in the exchange dominated approximation

For an infinite cylinder the variables in the Schrodinger equation separate and we can write

$$F_-(r,z) = m_0 J_p(k_\perp r) \Big[ A\cos\big(k_z(z-h/2)\big) + B\sin\big(k_z(z-h/2)\big) \Big] \tag{2.21}$$

yielding

$$\begin{pmatrix} m_x(\mathbf{r},t) \\ m_y(\mathbf{r},t) \end{pmatrix} = m_0 J_p(k_\perp r) \Big[ A\cos\big(k_z(z-h/2)\big) + B\sin\big(k_z(z-h/2)\big) \Big] \begin{pmatrix} \cos(p\phi - \omega t) \\ -\sin(p\phi - \omega t) \end{pmatrix}$$

. $\tag{2.22}$

Here $J_p$ is the Bessel function of order p. For a long but finite cylinder Eq. (2.21) should be a good approximation except for the cap modes.

If we take the modes +p and −p as degenerate we can superimpose them and form standing waves in $\varphi$, an operation we carry out in the next section. Inserting any of these forms into (2.8) yields the frequencies

$$\omega = \gamma \Big( H_0 + D_{ex}(k_z^2 + k_\perp^2) \Big). \tag{2.23}$$

If we adopt the boundary condition (discussed below)

$$\Big( \mathbf{n} \cdot \frac{d}{d\mathbf{r}} \Big) \mathbf{m} = 0 \tag{2.24}$$

where $\mathbf{n}$ is a vector normal to the surface, the values of $k_\perp$ will be fixed by the condition

$$\frac{dJ_p(k_\perp a)}{dk_\perp} = 0 \qquad (2.25)$$

where a is the cylinder radius. We write the solutions of Eq. (2.25) as $k_{p,n_r}$ where $n_r$ denotes the number of additional zeros of $J_p$ (other than those for $J_{p \neq 0}$ at r = 0) within the cylinder of radius a. We will find that (2.25) agrees quite well with the simulations.

For the finite cylinder we present the argument in two stages. In the first stage we assume that the inhomogenity in the static demagnetization field can be ignored and take $H(r,z) = H_0 - 4\pi N_\parallel M_0$ with $N_\parallel$ being the longitudinal demagnetization coefficient. The solution (2.21) continues to hold but the mode frequencies are given by

$$\omega(k_z, k_\perp) = \gamma\Big(H_0 - 4\pi N_\parallel M_0 + D_{ex}(k_z^2 + k_\perp^2)\Big). \qquad (2.26)$$

The allowed values of $k_\perp$ are given by $k_{p,n_r}$ as discussed above, but the quantization of $k_z$ is less simple. If the end caps are taken to be at $z = 0$ and $z = h$, then to have a definite parity under reflection in the mid plane at $z = h/2$, the mode function must depend on z as either $\cos(k_z(z - h/2))$ (even parity) or $\sin(k_z(z - h/2))$ (odd parity), but the association between $k_z$ and the parity depends on the boundary condition applied at the caps.

If the boundary condition is taken as $(\mathbf{n} \cdot \nabla)\mathbf{m} = 0$, then the allowed $k_z$ values are

$$k_z = \frac{\pi}{h}\upsilon_z, \quad \upsilon_z = 0, 1, 2 \cdots. \qquad (2.27)$$

Even parity is associated with even $\upsilon_z$ and odd parity with odd $\upsilon_z$.

If instead the boundary condition is taken as $\mathbf{m} = 0$, the allowed values of $k_z$ are

$$k_z = \frac{\pi}{h}\upsilon_z, \quad \upsilon_z = 1, 2 \cdots \qquad (2.28)$$

Now even parity is associated with odd $\upsilon_z$ and odd parity with even $\upsilon_z$.

The boundary condition obeyed by OOMMF mode functions is closer to $(\mathbf{n} \cdot \nabla)\mathbf{m} = 0$ than to $\mathbf{m} = 0$. In addition, they do have definite parity. Except for the two lowest frequency modes, which we call "cap modes" and which require a separate discussion, they are well fit by the $\cos(k_z(z - h/2))$ and $\sin(k_z(z - h/2))$ forms. However, it is advantageous to allow for a shift and write

$$k_z = \frac{\pi}{h} \upsilon'_z \qquad (2.29)$$

where

$$\upsilon'_z = \upsilon_z + \delta\upsilon_z \quad (\upsilon_z \geq 2) \qquad (2.30)$$

We can refer to $\delta\upsilon_z$ as an "end defect" analogous to the concept of a quantum defect in atomic spectroscopy[47]. With this correction Eq. (2.26) continues to be a good approximation to the mode frequencies.

We comment further on the boundary condition (2.24) that the normal derivative vanishes at the surface. The isotropic continuum exchange field $-D_{ex}\nabla^2 M$ arises from a microscopic $S_i \cdot S_j$ Heisenberg interaction, which has the property that for any pair of spins, the torque on $S_i$ due to $S_j$ cancels that on $S_j$ due to $S_i$. Thus, the total exchange torque on the body vanishes and Eq. (2.24) is the continuum expression of this fact. This argument dates back to Ament and Rado[48] and has been used by many authors since. Aharoni[49] offers a different derivation. Thus, it would appear to be very general, and valid for any $D_{ex}$, however small. For the magnetostatic limit, $D_{ex} = 0$, there is however no such condition on $M$. Turning on $D_{ex}$ perturbatively would then appear to lead to a contradiction. This is not so for the following reason.

In the boundary value problem for the spatial form of the eigenmodes, $D_{ex}$ multiplies the highest derivative, and is thus a singular perturbation from the mathematical point of view. Such perturbations are known to lead to thin boundary layers where the solution changes character rapidly[50]. Thus, while the normal derivative at the surface may formally be zero, there could be large curvature in the boundary layer, and the derivative of $m$ as we approach this layer need not be small. This is especially relevant for our OOMMF simulations, where the discretization into cells may: (a) be too coarse to reveal any boundary layer behavior, and (b) fundamentally preclude measurements of this derivative by fitting to the mode functions. In this case adoption of an end defect $\delta v_z$ is an effective practical procedure.

## 2.4. A variational solution for a cylinder of finite length

In the second stage of our argument, we attempt to include the inhomogeneity in the static demagnetizing field by adopting the trial form

$$m_{\pm}(\mathbf{r}) = Z(z)\, J_p(k_\perp r)\cos(p\varphi)\,. \qquad (2.31)$$

If we substitute this form in Eq. (2.17), together with some radially averaged z-dependent magnetic field $\bar{H}(z)$, we obtain the following one-dimensional eigenvalue problem

$$\omega Z(z) - \gamma \left[ \bar{H}(z) - D_{ex} \left( \frac{d^2}{dx^2} - k_\perp^2 \right) \right] Z(z) = 0 \qquad (2.32)$$

with the boundary conditions

$$Z'(0) = Z'(h) = 0 \,. \qquad (2.33)$$

In the spirit of this variational approach we could obtain $\bar{H}(z)$) by averaging with respect to $J_p^2(k_\perp r)$, which would lead to slight differences between modes with differing $k_\perp$. Alternatively, we can use the analytic expression for the dipole field along the cylinder axis $H_z(z, r = 0)$ that arises from spins which are fully aligned (as expected for the case where the exchange is totally dominant)[51]. Assuming a cylinder of radius a and height h, and setting z = 0 and z = h at the caps, the resulting demagnetization field along z-axis is

$$H_{demag}\left( r = 0, z \right) = -2\pi M_0 \left[ -\frac{h - z}{\sqrt{(h - z)^2 + a^2}} - \frac{z}{\sqrt{z^2 + a^2}} + 2 \right]. \qquad (2.34)$$

In the limit of $a/h \to 0$, $H_{demag}\left( r = 0, z \right) = 0$ (corresponding to an infinite rod) and in the limit of $a/h \to \infty$, $H_{demag}\left( r = 0, z \right) = -4\pi M_0$ (corresponding to a thin disk). Fig. 2.2 shows the resulting magnetic field for YIG cylinders having a diameter of 75nm and lengths of 75, 150, 300, 600, and 1200nm as calculated from Eq. (2.34) (dashed lines) and along the r = 0 axis by OOMMF. The close correspondence arises from the dominance of exchange in these small diameter samples.

## 2.5. Zeros of mode functions and mode labels

The demagnetizing field plays the role of an external potential in the Schrodinger equation (2.32), and the strong decrease in this field near the end caps leads to surface bound states or cap states whose wave functions die off exponentially away from the caps. In principle there could be many bound states, but for our parameters we find only one state at each cap. All higher energy states are extended along the z direction, and since the demagnetization field is essentially uniform in the bulk of the cylinder, their wavefunctions behave approximately as sinusoidal standing waves.

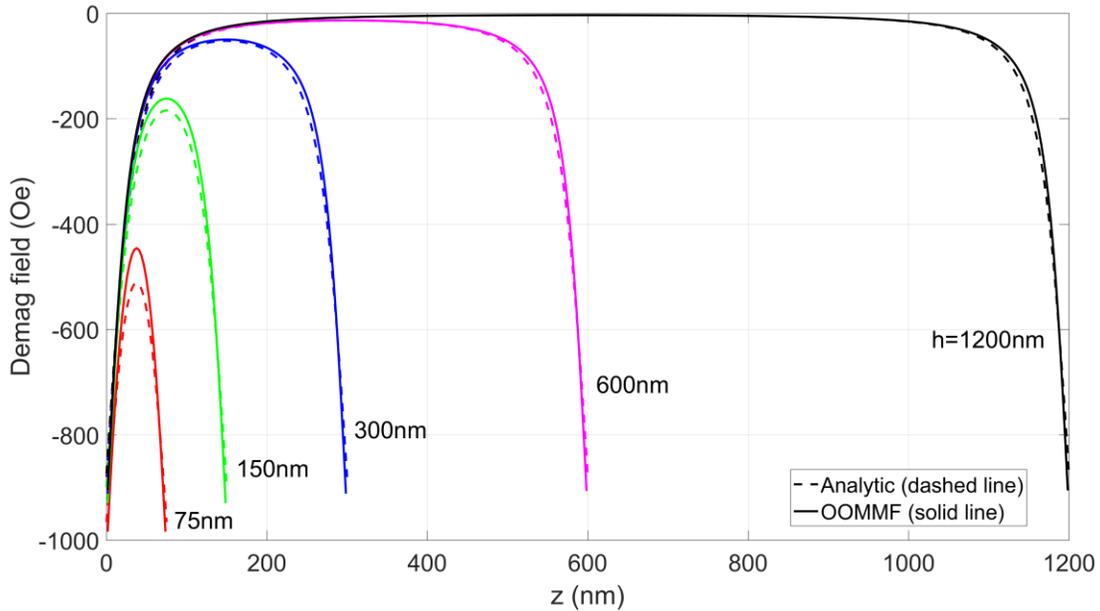

**Fig. 2.2. The dashed line shows the analytic demagnetization field calculated from Eq. (2.32) using $4\pi M_0 = 1750 \, Oe$ for YIG cylinders with a diameter of 75nm and five different lengths of 75, 150, 300, 600, 1200nm in a field of 2000Oe. For comparison the solid line shows the field computed by OOMMF along the line r = 0. Note how for long cylinders the field profile near one cap is insensitive to the presence of the other cap.**

Let us now recall that for a one-dimensional Schrodinger equation with a reflection symmetric potential, the states with successively higher energy alternate in parity and have successively increasing number of zeros, with the lowest energy state having no nodes and even parity. Further their wave functions must be mutually orthogonal. For our problem, these theorems are satisfied as follows. The two cap states are nearly degenerate but they are admixed by tunneling to form even and odd parity states with zero and one node respectively. (See, however, the discussion in Sec. 7 on how the inclusion of dipole-dipole interactions modifies the energy ordering.) The first extended state must then have even parity and two nodes. To be approximated by $\cos\big(k_z(z - h/2)\big)$ and to be orthogonal to the cap states, we must have $k_z \lesssim 2\pi / h$ corresponding to $\nu_z = 2$ and a negative end defect $\delta\nu_z$. Higher extended states must have higher values of $\nu_z$. In this way we see the need for the restriction $\nu_z \geq 2$ in Eq. (2.30) and for the end defect at the same time.

For each value of p and $n_r$, we could label the differently quantized modes along z by the number of their zeros. The lowest extended or bulk mode in any family with given p and $n_r$ would then have the label $\nu_z = 2$, while the cap mode would be labeled $\nu_z = 0$. This is unaesthetic and also does not differentiate between the physically different character of the cap modes vis a vis the bulk modes. We therefore label the bulk modes by an index $n_z$, with

$$n_z = \nu_z - 2 \qquad (2.35)$$

For the cap modes we replace the number $n_z$ by the letters 'g' (gerade, even parity) and 'u' (ungerade, odd parity). For reference we summarize the correspondence between the number of zeros and the mode labels as follows:

| No. of zeros | Mode label |
|:---:|:---:|
| 0 | g |
| 1 | u |
| $\nu_z$ | $n_z = \nu_z - 2$ |

The modes are labeled by the scheme $(p\,n_r\,n_z)$ with the letters 'g' or 'u' for cap modes in lieu of $n_z$. In particular the mode nominally identified as the uniform FMR mode has the label (000) (but see the discussion in Sec. 7).

2.6. Numerical results for the variational approximation and comparison with simulations

We now describe some results from the numerical integration of Eq. (2.32) together with the position dependence of $H_{demag}(r = 0, z)$ given by Eq. (2.34). Imposing the boundary condition (2.33) at the faces then yields $Z(z)$ together with the eigenvalues $\omega = \omega(n_z(p, n_r))$, where for the general case $n_z(p, n_r)$ denotes the eigenvalue for given values of the azimuthal and radial mode numbers, $p$ and $n_r$. Given that we have neglected the transverse dipolar field in obtaining Eq. (2.32) we expect the resulting eigenvalues to be most accurate in the limit of large $n_z$ mode numbers, and particularly when both $p = 0$ and $n_r = 0$ (which corresponds to $\theta = 0$ in Eq. (1.3)).

The dashed lines in figure 2.3 show the resulting form of $Z(z)$ for the lowest lying cap mode with p = 0 and no radial nodes for cylinders with a diameter of 75nm and heights of 75, 150, 300, 600, and 1200 nm in a field of 2000G. Note the approximately exponential decay of the amplitude as we proceed deep into the interior for the longer samples confirming their surface like character. Also shown are the OOMMF simulations obtained using procedures to be outlined below (the fact that their amplitudes do not go strictly to zero in longer samples arises from a contamination from other modes). Accompanying antisymmetric modes (not shown) are also highly localized while in addition having a node at the cylinder midpoint.

As is evident, the semi-analytic results for $Z(z)$ are surprisingly good for $h \geq 300\,\text{nm}$. The frequencies however are not. These could be improved by including the transverse dipolar field using perturbation theory, which will raise the frequency. We have not attempted this exercise since our approximate treatment is quite rough in the first place and it would not add to our qualitative understanding.

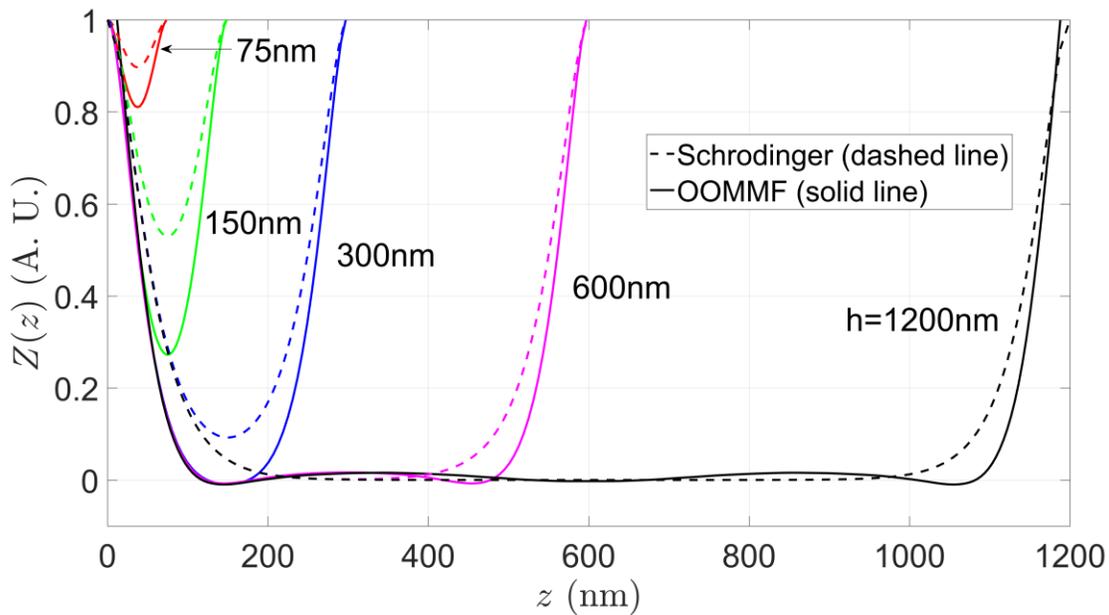

**Fig. 2.3. The dashed lines show the behavior of the mode function Z(z) vs. z obtained from integrating Eq. (2.32) for the (00g) cap mode for cylinders with a diameter of 75nm and heights of 75, 150, 300, 600, and 1200nm. The solid lines show the OOMMF simulation results for the same parameters.**

As a crude estimate of the cap mode frequency we can compare it with the frequency of a hat box (disc) with a radius equal to its height. Reported demagnetization coefficients[52] for this aspect ratio are $N_{\parallel} = 0.4745$ and $N_{\perp} = 0.2628$. For a field $H_0 = 2.000\,\text{kOe}$ and $4\pi M_0 = 1.750\,\text{kOe}$ Eq. (1.2) yields f = 4.568 GHz. For our exchange dominated sample it is reasonable to add a correction of order $\gamma D_{ex} / a^2 = 1.0\,\text{kOe}$ which raises the frequency to 5.6 GHz which is to be compared with the OOMMF value of 6.64 GHz.

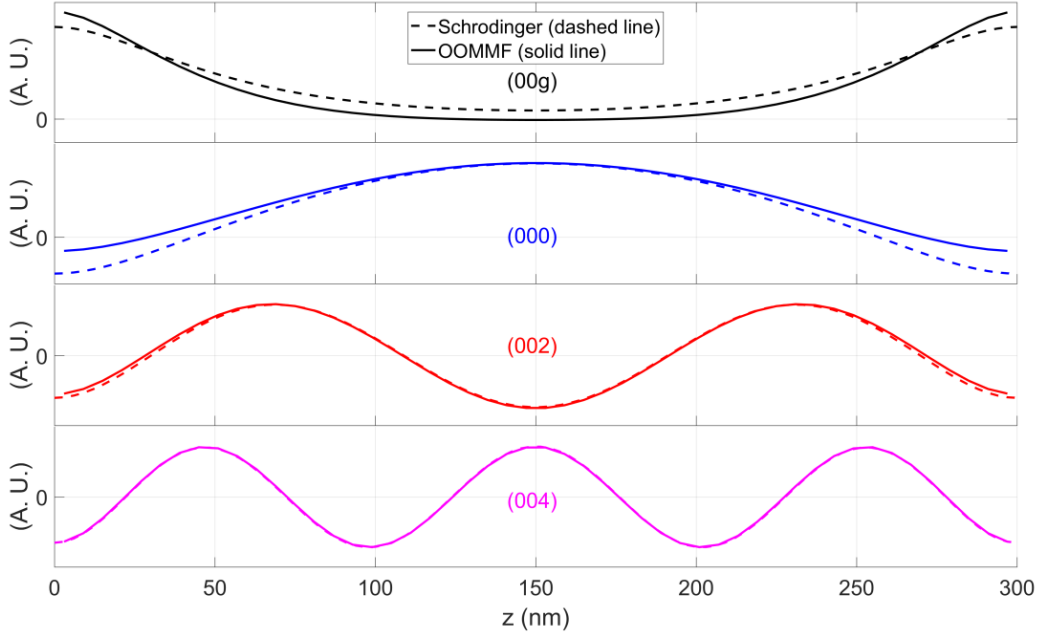

**Fig. 2.4. The dashed line shows the mode function Z(z) vs. z obtained from integrating Eq. (2.35) for the (00g), (000), (002), and (004) modes. The solid line shows the corresponding forms arising from OOMMF. Note the behavior at the cylinder faces closely conforms with the boundary condition (2.24).**

| Table I. End defect for (00n) modes. | | | |
|---|---|---|---|
| Mode Label | $\nu_z$ | $hk_{z,fit}/\pi$ | $\Delta\nu_z$ |
| 000 | 2 | 1.62 | $-0.38$ |
| 001 | 3 | 2.50 | $-0.50$ |
| 002 | 4 | 3.68 | $-0.32$ |
| 003 | 5 | 4.78 | $-0.22$ |
| 004 | 6 | 5.84 | $-0.16$ |
| 005 | 7 | 6.96 | $-0.04$ |
| 006 | 8 | 7.90 | $-0.10$ |
| 007 | 9 | 8.92 | $-0.08$ |
| 008 | 10 | 10.01 | $+0.01$ |
| 009 | 11 | 10.91 | $-0.09$ |
| 00,11 | 13 | 12.96 | $-0.04$ |
| 00,13 | 15 | 14.98 | $-0.02$ |
| 00,15 | 17 | 16.92 | $-0.08$ |

Calculations for the extended states with mode number $n_z = 0, 1, \cdots$ were also performed. Here we encounter progressively higher mode frequencies scaling approximately as $n_z^2$. Figure 2.4 below shows the result of such calculations for the (00g), (000), (002), and (004) modes. Table I lists values of $k_z h/\pi$, $\nu_z$ and $\delta\nu_z$ for these and neighboring modes. Note that $\delta\nu_z \rightarrow 0$ with increasing $\nu_z$. We will explain why modes (00,10), (00,12), and (00,14) are not in this table at the end of Sec. 3.2.

## 3. Computational Approaches

The material studied here is YIG which was chosen for its long mode lifetimes. Whether these long lifetimes survive in submicron structures is an open question. The majority of the studies were for a sample with h = 4d = 300 nm. The material parameters used are typical for YIG [53]: $\gamma = 2\pi \times 2.8$ GHz/kOe, saturation magnetization $M_s = 139$ emu/cm$^3$, damping constant $\alpha = 5 \times 10^{-5}$, and exchange constant $A_{ex} = 3.5 \times 10^{-7}$ erg/cm. The applied field was 2 kOe along z direction. Damping was turned on to relax the system to its initial state, and turned off after the system was excited for most simulations. In the few that it was not, it was too small to have any significant effect.

As noted above our simulations were carried out with the OOMMF code developed by the U. S. National Institute for Standards and Technology. This program divides a chosen sample into cells on a rectilinear grid and numerically integrates the LL equation in time for their magnetizations, $\mathbf{M}_i(t)$, as they evolve under the influence of the torques acting on them arising from an external field, the nearest neighbor exchange interaction, and the dipolar fields of the remaining cells (anisotropy fields can be included but will be ignored in what follows). Each magnetic moment is located at the center of the cell. The number of cells scales with the cube of a characteristic sample dimension but was nominally fixed at 1/54 cells/(nm)$^3$ corresponding to a cell size of $3 \, nm \times 3 \, nm \times 6 \, nm$ for the $d = 75 \, nm, h = 300 \, nm$ sample. There are 50 cells in the z direction, and 489 cells in the xy plane (489/625 = 0.7824 vs. $\pi/4 = 0.7854$). In Sec. 7, we simulate samples with other values of h. As described there, we then use cells with the same x and y dimensions (3 nm × 3 nm), but depending on the value of h, the dimension $\Delta z$ is adjusted appropriately.

### 3.1. Static equilibrium

Prior to exciting the system, the spins were initially aligned along the cylinder axis (parallel to the external field) after which the system was evolved in time (with damping) until it stabilizes in an equilibrium configuration. Various tests can be applied to determine that it is a global equilibrium state. This part of our simulations yields the static magnetic field distribution which could also be used for the calculations in Section 2.

## 3.2. Exciting the system

Several different excitation schemes were utilized. In the simplest of these, all spins were tipped by a small fixed angle relative to their equilibrium orientations in a plane containing the z axis as an initial condition. This favors the excitation of uniformly precessing modes. To drive a *particular* non-uniform mode the spins were tipped from their equilibrium positions in a manner *that mimics the mode* (such as that obtained as the mode pattern in a prior simulation)[54]. To drive a broader *spectrum of modes* that is localized around a time $t_0$ and some position $\mathbf{r}_0$ we tip the spins in some direction according to the function[41]

$$F(t, \mathbf{r}) = A \frac{\sin[\Delta\omega(t - t_0)]}{(t - t_0)} \frac{\sin[\Delta k_x(x - x_0)]}{(x - x_0)} \frac{\sin[\Delta k_y(y - y_0)]}{(y - y_0)} \frac{\sin[\Delta k_x(z - z_0)]}{(z - z_0)} \quad (3.1)$$

where $\Delta\omega$, $\Delta k_x$, $\Delta k_y$ and $\Delta k_x$ control the extent to which the excitation is localized in time and space. Here x, y, z denote cell coordinates. Such pulses can also be introduced at multiple times and positions to favor the excitation of modes with differing spatial and temporal properties. In particular inclusion of only the last factor induces modes propagating along z. Forms can be constructed that favor the excitation of radial or azimuthal modes. Finally, some simulations were performed in which individual spins were tipped in random directions within some specified average angular range. This excites a very broad range of modes and if the tipping angles are large (e.g., approaching 180º) generates a "hot" system, from which it is difficult to extract clear modes. Altogether, we tried more than ten different excitation pulses in an effort to identify modes with different symmetry and numbers of nodes. Despite this, our mode table (see Sec. 4) has gaps. In some cases, modes are nearly degenerate [for example modes (003), (-10g), and (-10u)] and cannot be easily resolved. In others, they were too high in frequency to be seen with the particular excitation pulse employed. We are confident that these modes exist and that our mode classification is complete.

## 3.3. Identifying modes

As the system state simulated by OOMMF evolves in time from some chosen initial configuration, the magnetization vectors $\mathbf{m}(\mathbf{r_i}, t)$ at the (discrete) cell sites $\mathbf{r_i}$ are recorded at regular time intervals. From this data set we can perform a cell by cell fast-Fourier transform (FFT) within some chosen time interval available from the simulation to obtain the complex quantities $\mathbf{m}(\mathbf{r_i}, \omega)$. We stress that the OOMMF simulation does not assume that $\mathbf{M_{eq}}(\mathbf{r})$ is along $\hat{\mathbf{z}}$ or that the deviations $\mathbf{m}(\mathbf{r_i}, t)$ are in the x-y plane, although the most useful information is contained in these components for low amplitude mode studies. From the FFT, we follow McMichael and Stiles[5] and construct the cell-wise power spectra,

$$S_x(\mathbf{r_i}, \omega) = |m_x(\mathbf{r_i}, \omega)|^2, \qquad (3.2)$$

together with their sum over the entire sample,

$$\bar{S}_x(\omega) = \sum_i S_x(\mathbf{r_i}, \omega), \qquad (3.3)$$

and likewise for $S_y(\mathbf{r_i}, \omega)$ and $\bar{S}_y(\omega)$. As noted by them, this definition of a power spectrum is very different from the power spectrum of the integrated magnetization (total magnetic moment of the sample), which is what makes them so useful in mode identification; in particular, the frequencies where these total sample power spectra have sharp maxima are identified as possible mode frequencies of the system. As an example, the power spectra in Eq. (3.3) are given in Fig. 3.1 which shows various modes including a very high frequency mode that is aliased to less than 5 GHz due to a Nyquist critical frequency of 50 GHz.

Suppose a mode has been identified at a frequency $\omega_a$. With the sign conventions used in our numerical FFT program, the mode pattern associated with this frequency is given by

$$\mathbf{m}^{(a)}(\mathbf{r_i}, t) = \Re\left(\mathbf{m}(\mathbf{r_i}, \omega)e^{i(\omega_a t + \phi)}\right). \qquad (3.4)$$

The phase $\phi$ is arbitrary and amounts to a choice of the zero of time. To avoid unnecessary minus signs, we choose $\phi = 3\pi/2$ which gives

$$\mathbf{m}^{(a)}(\mathbf{r_i}, 0) = \Im\mathbf{m}(\mathbf{r_i}, \omega_a), \qquad \mathbf{m}^{(a)}(\mathbf{r_i}, T/4) = \Re\mathbf{m}(\mathbf{r_i}, \omega_a), \qquad (3.5)$$

with $T = 2\pi/\omega_a$ being the time period of the mode. Hence, by examining the imaginary and real parts of the vector $\mathbf{m}(\mathbf{r_i}, \omega_a)$ we can, respectively, obtain the real-space vector magnetization at some time and a quarter cycle later for the spatial pattern associated with some nominal mode at the specific frequency. By plotting these vector fields, we can get a highly visual depiction of the mode, permitting easy mode assignment and further analysis.

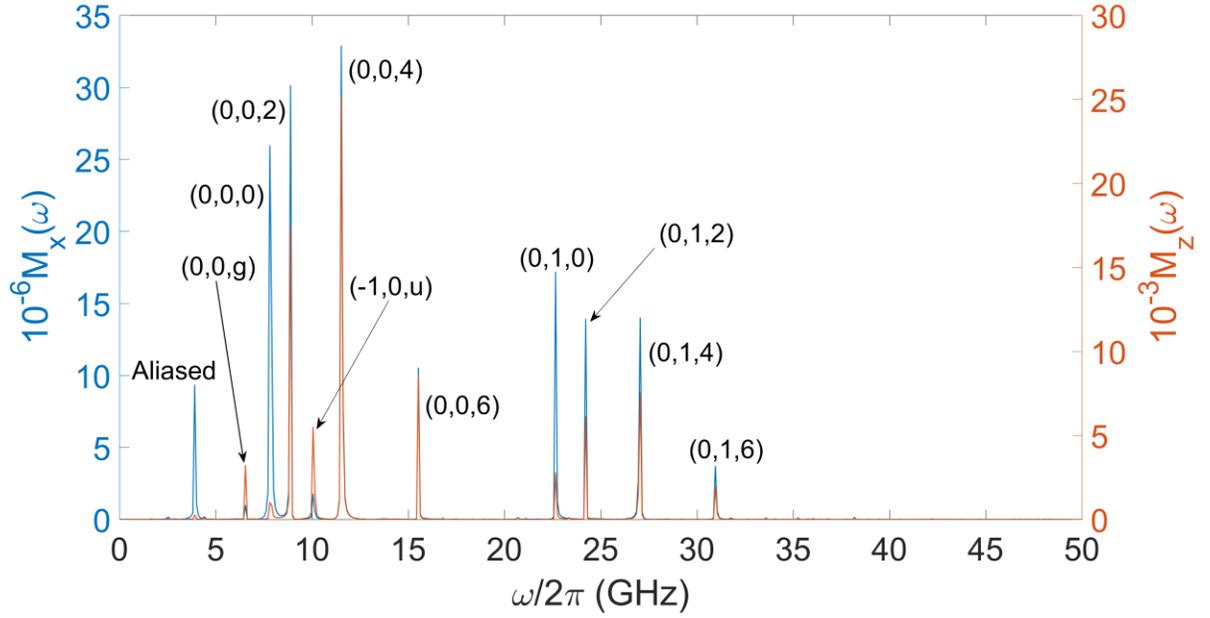

**Fig. 3.1. An example of an FFT spectrum. The left and right y axes show** $2\overline{S}_x(\omega)\big/ N_{sample}^2$ **and** $2\overline{S}_z(\omega)\big/ N_{sample}^2$ **. See Eq. (3.3). A broad sinc pulse as described in Eq. (3.1) was used to excite this spectrum.**

In constructing the power spectrum, modes with higher frequencies than the inverse of the chosen integration time step, which violate the Nyquist sampling criterion and are then aliased to lower frequencies, must be identified and rejected. In the exchange dominated samples considered here some of them can be identified as spurious peaks with frequencies lower than the known uniform modes, but in dipole dominated larger samples genuine modes below the uniform modes are expected. More generally they must be identified by altering the time interval over which the transform is performed to determine if some mode moves its position. Most simulations were done with a time step of 10 ps and for a duration of 10.24 ns.

A curious spatial aliasing was also observed (as evidenced by a rapid spatial variation of the mode intensity on the scale of the cell period) in some patterns; it is thought to be associated with a spatial FFT that is performed to calculate the dipole field in the underlying program. Such modes must also be rejected.

### 3.4. Implications of the ±p degeneracy for OOMMF patterns

Using the procedures described we can construct mode maps in chosen planes by plotting the complex cell amplitudes, $\mathbf{m}(\mathbf{r},\omega) = m_x(\mathbf{r},\omega)\hat{\mathbf{x}} + m_y(\mathbf{r},\omega)\hat{\mathbf{y}}$, at frequencies where the power spectrum shows maxima. On the basis of these patterns we are typically able to assign approximate mode numbers p, $n_r$ and $n_z$, and designate them as $\mathbf{m}^{(pn_rn_z)}(\mathbf{r}, \omega^{(pn_rn_z)})$ for that frequency, $\omega = \omega^{(pn_rn_z)}$. The p mode number requires special attention as we now discuss. In what immediately follows we will drop the mode designation, regarding it as being understood.

Writing the complex function $\mathbf{m}(\mathbf{r}, \omega)$ in component form as

$$\mathbf{m}(\mathbf{r}, \omega) = \begin{pmatrix} m_x(\mathbf{r}, \omega) \\ m_y(\mathbf{r}, \omega) \end{pmatrix} = \begin{pmatrix} m_x'(\mathbf{r}, \omega) + i m_x''(\mathbf{r}, \omega) \\ m_y'(\mathbf{r}, \omega) + i m_y''(\mathbf{r}, \omega) \end{pmatrix}, \tag{3.6}$$

the corresponding behavior in the time domain follows as

$$\mathbf{m}(\mathbf{r}, t) = \begin{pmatrix} m_x(\mathbf{r}, t) \\ m_y(\mathbf{r}, t) \end{pmatrix} = \Re \begin{pmatrix} m_x'(\mathbf{r}, \omega) + i m_x''(\mathbf{r}, \omega) \\ m_y'(\mathbf{r}, \omega) + i m_y''(\mathbf{r}, \omega) \end{pmatrix} e^{-i\omega t}$$

$$= \begin{pmatrix} m_x'(\mathbf{r}, \omega)\cos\omega t + m_x''(\mathbf{r}, \omega)\sin\omega t \\ m_y'(\mathbf{r}, \omega)\cos\omega t + m_y''(\mathbf{r}, \omega)\sin\omega t \end{pmatrix}. \tag{3.7}$$

If the modes of the system have a pure p character, as in Eq. (2.11), we can write the above components as

$$m_x'(\mathbf{r}, \omega) = m(r, z, p, \omega)\cos(p\varphi), \qquad m_x''(\mathbf{r}, \omega) = m(r, z, p, \omega)\sin(p\varphi), \tag{3.8a, b}$$

$$m_y'(\mathbf{r}, \omega) = -m(r, z, p, \omega)\sin(p\varphi), \qquad m_y''(\mathbf{r}, \omega) = m(r, z, p, \omega)\cos(p\varphi). \tag{3.8c, d}$$

Hence, we can write $\mathbf{m}(\mathbf{r}, t)$ as

$$\mathbf{m}(\mathbf{r}, t) = \begin{pmatrix} m_x(r, z, p, t) \\ m_y(r, z, p, t) \end{pmatrix} = m(r, z, p, \omega) \begin{pmatrix} \cos(p\varphi - \omega t) \\ -\sin(p\varphi - \omega t) \end{pmatrix}. \tag{3.9}$$

Here the magnetization vector rotates as $\varphi$ changes with a *radially symmetric* amplitude. In our approximation, where the variables r and z separate, we would write the solutions that have even parity as

$$\begin{pmatrix} m_x(\mathbf{r}, t) \\ m_y(\mathbf{r}, t) \end{pmatrix} = m_0 \cos\big(k_z(z - h/2)\big) J_p(k_\perp r) \begin{pmatrix} \cos(p\varphi - \omega t) \\ -\sin(p\varphi - \omega t) \end{pmatrix}. \tag{3.10}$$

If the modes of the system have a pure p character and in addition p and $-p$ are degenerate we can form symmetric and antisymmetric standing wave super-positions of the two forms of Eq. (3.10) to obtain

$$\mathbf{m}(\mathbf{r},t) = \begin{pmatrix} m_x(r,z,p,t) \pm m_x(r,z,-p,t) \\ m_y(r,z,p,t) \pm m_x(r,z,-p,t) \end{pmatrix}$$

$$= m(r,z,p,\omega) \begin{pmatrix} \cos(p\varphi - \omega t) \pm \cos(-p\varphi - \omega t) \\ -\sin(p\varphi - \omega t) \mp \sin(-p\varphi - \omega t) \end{pmatrix} \tag{3.11}$$

which results in the following two forms

$$m(r,z,p,\omega)\cos(p\varphi) \begin{pmatrix} \cos(\omega t) \\ \sin(\omega t) \end{pmatrix} \tag{3.12a}$$

$$m(r,z,p,\omega)\sin(p\varphi) \begin{pmatrix} \sin(\omega t) \\ -\cos(\omega t) \end{pmatrix}, \tag{3.12b}$$

or if our model product form is assumed,

$$m_0 \cos\big(k_z(z - h/2)\big) J_p(k_\perp r)\cos(p\varphi) \begin{pmatrix} \cos(\omega t) \\ \sin(\omega t) \end{pmatrix}, \tag{3.13a}$$

$$m_0 \cos\big(k_z(z - h/2)\big) J_p(k_\perp r)\sin(p\varphi) \begin{pmatrix} \sin(\omega t) \\ -\cos(\omega t) \end{pmatrix}. \tag{3.13b}$$

Here again the spin direction rotates (in both senses) but now the amplitude is modulated in $\varphi$. A similar discussion applies to the solutions that are odd parity.

In Appendix B we discuss how a pure p mode pattern can be extracted from a superposition of +p and −p OOMMF patterns.

### 3.5. Separating nearly degenerate modes

As noted above, in the absence of the dipole interaction modes with +p and −p are degenerate. When this interaction is present such modes are split; i.e., according to the discussion of section 3.4 our eigen modes will be running waves in $\varphi$. But the splitting rapidly decreases for larger mode numbers, and for running times less than $t \leq \Delta\omega^{-1}$, where $\Delta\omega$ is the splitting, the power spectrum displays a single (slightly broadened) peak at the mode frequency. In order to resolve the splitting in a power spectrum the OOMMF run times must be increased.

When the splitting is not resolved in the power spectrum the resulting mode patterns display a standing wave character. For a few of these we used the standing wave mode pattern as an initial configuration and ran the program long enough to display a beat pattern in time from which the splitting could be accurately determined. An example of this technique is shown in Fig. 3.2 for the cylinder with d = 75 nm and h = 300 nm. Fig. 3.2a shows the case of the (±105) modes with an average frequency of 17.29 GHz. Note a beat waist occurs at t = 32.68 ns from which we calculate the mode splitting as 30.60 MHz.

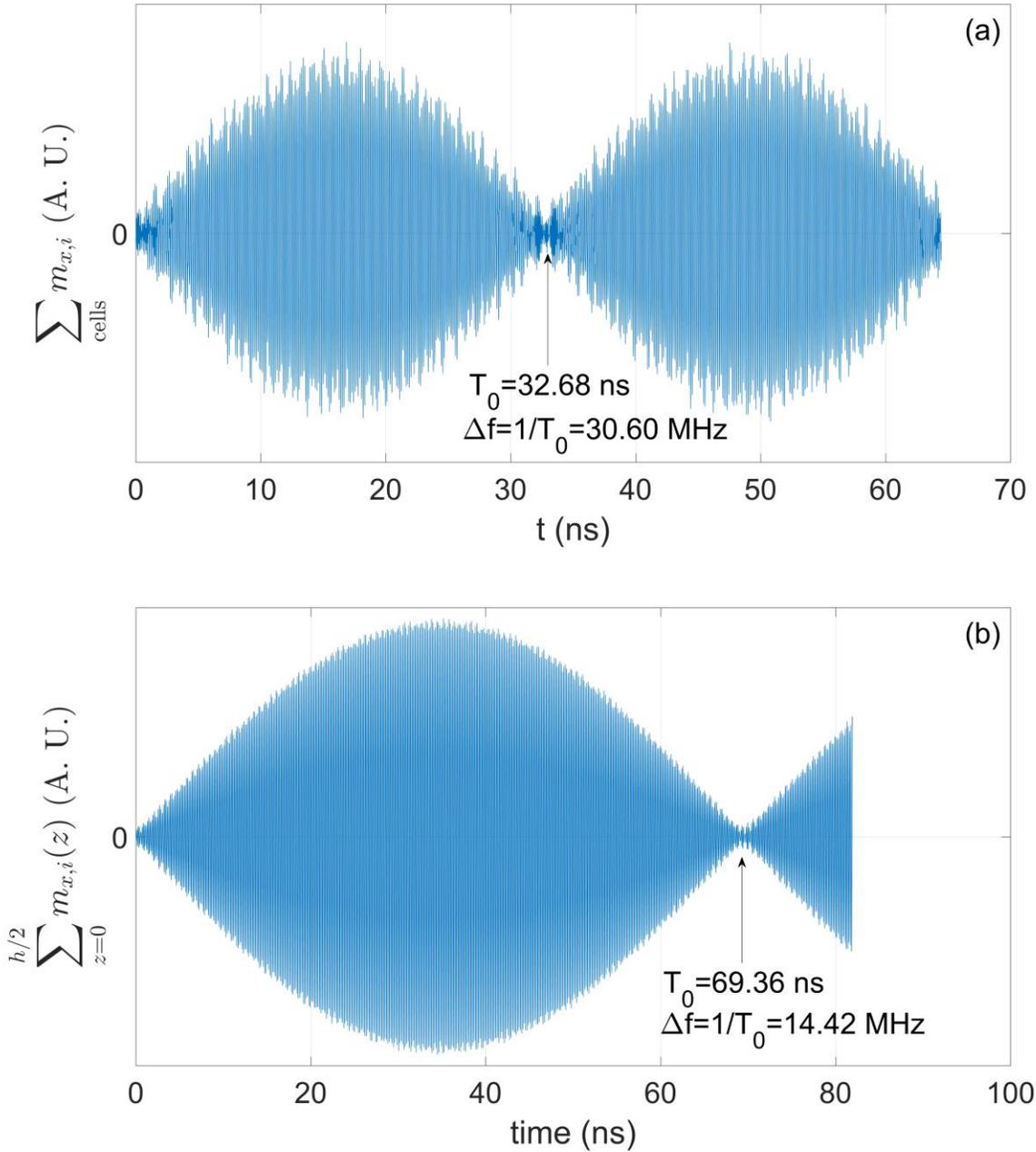

Fig. 3.2. Beat patterns emerging from the time evolution arising from the splitting of (a) the (±105) modes with an average frequency 17.29 GHz and a splitting of 30.60 MHz, (b) the p = 0, $n_r$ = 0, g and u cap modes with an average frequency 6.54 GHz and a splitting of 14.4MHz.

A very small splitting also occurs between the symmetric (or gerade, denoted g) and anti-symmetric (ungerade, denoted u) combinations of the cap modes. In the Schrödinger equation language of Sec. 2, this is a tunnel splitting between the surface bound states. This splitting is intrinsically small and hard to resolve in long cylinders, although we have resolved it for the $p = 0$, $n_r = 0$, g and u cap modes for the cylinder with h = 300 nm. Now $f = 6.54$ GHz and $\Delta f = 14.4$ MHz. The corresponding beat pattern is shown in Fig. 3.2b.

## 4. Frequencies of low-lying modes

Most computations were carried out on a YIG sample with $h = 4d = 8a = 300$ nm in a static field of $H_0 = 2000$ Oe. To test the behavior at small and large $k_z$ some calculations were carried out for samples with h = 7.5, 37.5, 75, 150, 600 and 1200 nm. The material parameters used are typical for YIG, as given earlier in section 3.

Table II lists the frequencies $f_{p n_r n_z} = \omega_{p n_r n_z} / 2\pi$ of low lying modes as obtained from the peaks in the power spectrum; all entries are for $h = 4d = 8a = 300$ nm and in a static field of $H_0 = 2000$ Oe. The mode numbers come from a comparison of the accompanying mode pattern with the forms discussed in Sec. 2 with special attention to the number of radial and longitudinal zeros and how **m** winds around the z axis. By fitting to the form (2.21) we can assign discrete wavevectors, $k_{\perp, n_r}$ and $k_{n_z}$; these values are used for a comparison with the HK formula as we describe in the next section. All modes with $p \neq 0$ have a node at $r = 0$; for larger values of $k_\perp$, additional radial nodes can be present and their mode number is denoted as $n_r \neq 0$. Modes listed as $\pm p$ are nearly degenerate in the sense discussed above, and the mode patterns display standing waves in $\varphi$ as described by Eq. (3.9a,b).

Table II. The mode frequencies for a YIG cylinder with a height of 300nm and a diameter of 75nm in a static magnetic field $H_0 = 2000$ Oe, organized into families with given mode numbers p and $n_r$. Multiply listed modes (e.g., (–115) and (± 115)) were observed with both pure and with mixed +p and –p character, depending on the methodology used to extract them (e.g. FFT vs. beat pattern).

## Table II

## Mode Frequencies vs. p, $n_r$, and $n_z$

| p | $n_r$ | $n_z$ | f (GHz) | p | $n_r$ | $n_z$ | f (GHz) | p | $n_r$ | $n_z$ | f (GHz) |
|---|---|---|---|---|---|---|---|---|---|---|---|
| 0 | 0 | Even cap (g) | 6.543 | 1 | 0 | g | 10.35 | ±1 | 0 | 5 | 17.29 |
| 0 | 0 | Odd cap (u) | 6.543 | 1 | 0 | u | 10.35 | ±1 | 0 | 7 | 21.78 |
| 0 | 0 | 0 | 7.813 | 1 | 0 | 1 | 11.91 | ±1 | 0 | 11 | 33.98 |
| 0 | 0 | 1 | 8.105 | 1 | 0 | 3 | 13.96 | ±1 | 1 | g | 35.16 |
| 0 | 0 | 2 | 8.887 | 1 | 0 | 5 | 17.29 | ±1 | 1 | u | 35.16 |
| 0 | 0 | 3 | 10.06 | 1 | 0 | 7 | 21.78 | ±1 | 1 | 1 | 36.91 |
| 0 | 0 | 4 | 11.52 | 1 | 0 | 9 | 27.44 | ±1 | 1 | 3 | 39.26 |
| 0 | 0 | 5 | 13.38 | 1 | 0 | 11 | 33.98 | ±1 | 1 | 5 | 42.68 |
| 0 | 0 | 6 | 15.53 | 1 | 0 | 13 | 41.5 | ±1 | 1 | 9 | 52.83 |
| 0 | 0 | 7 | 17.97 | 0 | 1 | 0 | 22.66 | ±2 | 0 | g | 15.82 |
| 0 | 0 | 8 | 20.7 | 0 | 1 | 1 | 23.24 | ±2 | 0 | u | 15.82 |
| 0 | 0 | 9 | 23.63 | 0 | 1 | 2 | 24.22 | ±2 | 0 | 1 | 18.07 |
| 0 | 0 | 11 | 30.27 | 0 | 1 | 3 | 25.49 | ±2 | 0 | 2 | 18.36 |
| 0 | 0 | 13 | 37.79 | 0 | 1 | 4 | 27.05 | ±2 | 0 | 3 | 19.63 |
| 0 | 0 | 15 | 46.09 | 0 | 1 | 5 | 28.91 | ±2 | 0 | 5 | 22.95 |
| −1 | 0 | g | 10.06 | 0 | 1 | 6 | 30.96 | ±2 | 0 | 9 | 33.11 |
| −1 | 0 | u | 10.06 | 0 | 1 | 7 | 33.4 | ±2 | 0 | 11 | 39.65 |
| −1 | 0 | 1 | 11.72 | 0 | 1 | 8 | 36.04 | ±2 | 0 | 15 | 55.37 |
| −1 | 0 | 2 | 12.7 | 0 | 1 | 9 | 38.96 | ±3 | 0 | g | 24.02 |
| −1 | 0 | 3 | 13.87 | 0 | 1 | 10 | 42.19 | ±3 | 0 | u | 24.02 |
| −1 | 0 | 4 | 15.43 | 0 | 1 | 11 | 45.51 | ±3 | 0 | 1 | 25.59 |
| −1 | 0 | 5 | 17.19 | 0 | 1 | 12 | 49.12 | ±3 | 0 | 3 | 27.83 |
| −1 | 0 | 6 | 19.34 | −1 | 1 | g | 35.16 | ±3 | 0 | 5 | 31.25 |
| −1 | 0 | 7 | 21.78 | −1 | 1 | u | 35.16 | ±3 | 0 | 7 | 35.84 |
| −1 | 0 | 8 | 24.41 | −1 | 1 | 3 | 39.26 | ±3 | 0 | 11 | 47.95 |
| −1 | 0 | 9 | 27.34 | −1 | 1 | 5 | 42.68 | | | | |
| −1 | 0 | 10 | 30.57 | −1 | 1 | 6 | 44.82 | | | | |
| −1 | 0 | 11 | 33.98 | 0 | 2 | g | 55.57 | | | | |
| −1 | 0 | 12 | 37.6 | 0 | 2 | u | 56.64 | | | | |
| −1 | 0 | 13 | 41.5 | 0 | 2 | 2 | 58.3 | | | | |
| −1 | 0 | 14 | 45.51 | 0 | 2 | 4 | 61.23 | | | | |
| −1 | 0 | 15 | 49.71 | 0 | 2 | 6 | 65.53 | | | | |
| | | | | 0 | 2 | 8 | 70.41 | | | | |

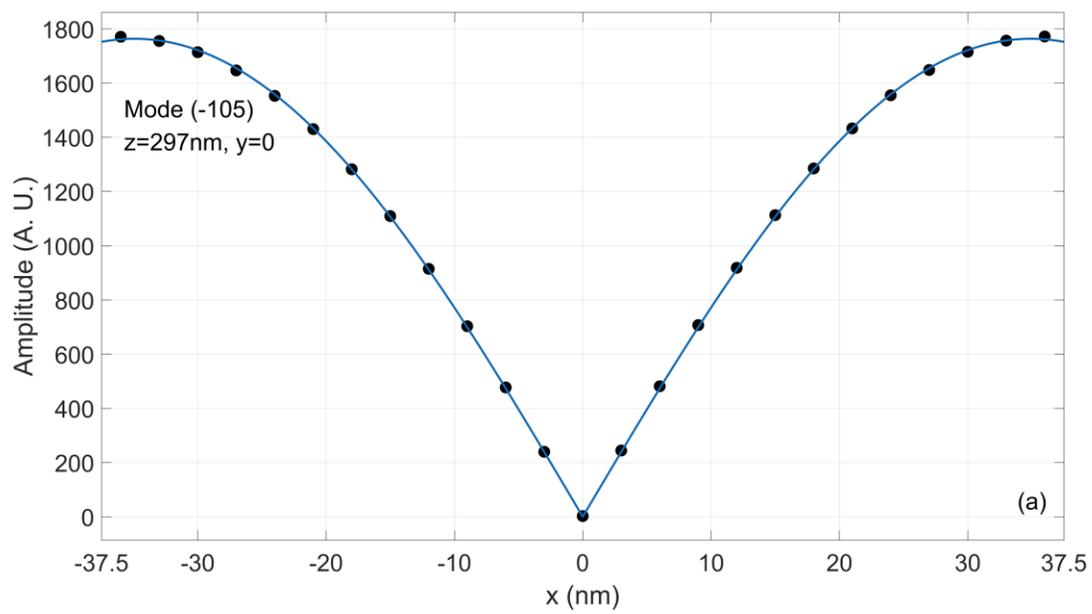

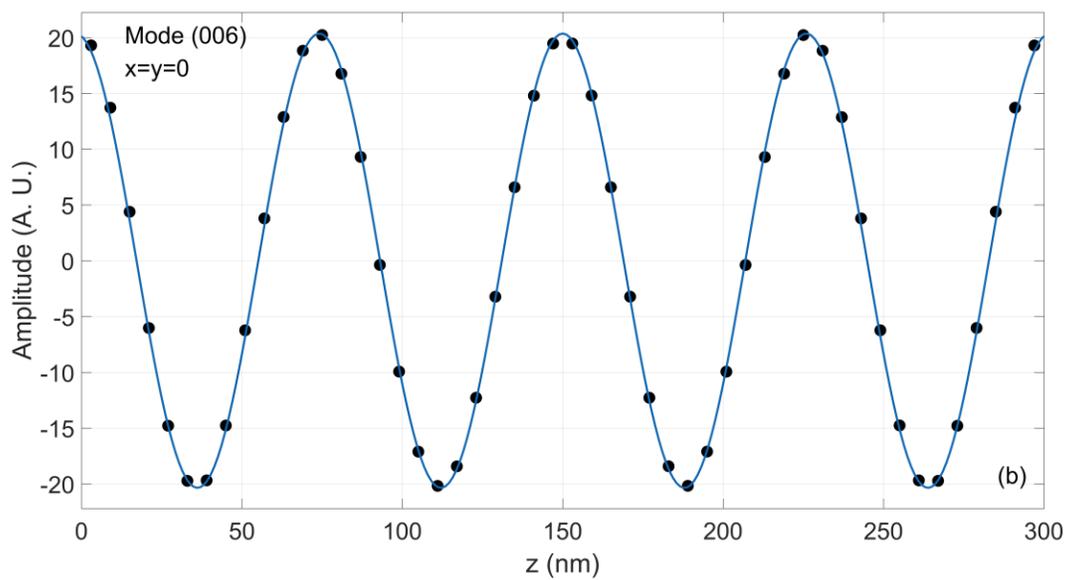

**Fig. 5.1. a) Typical fit of the radial OOMMF amplitude to the function $J_1(k_\perp r)$ for the $(-105)$ mode. b) Typical fit of the axial OOMMF amplitude to the function $\cos k_z(z - h/2)$ for the $(006)$ mode.**

# 5. Comparison of simulated frequencies with the Herring-Kittel expression

It is interesting to examine the extent to which the OOMMF mode frequencies in Table II can be represented by the Herring-Kittel frequencies as given by Eq, (1.3). To do this we need values of $k_z$ and $k_{p,n_r}$ for the extended modes. A preliminary value of $k_z$ follows from counting the number of nodes along z. Better values emerge[55] from fitting $m_r(z, r = 0, \omega)$ to $\cos(k_z(z - h/2))$ or $\sin(k_z(z - h/2))$. Values for $k_{p,n}$ were obtained by fitting $m_r(z, r, \omega)$ to $J_p(k_{p,n}r)$ at some z with $k_{p,n}$ as an adjustable parameter. Fig. 5.1a, b shows examples of such fits. Note that although we do not employ it to find $k_z$ and $k_\perp$, the boundary condition at the faces for this mode closely approximates maximum amplitude as opposed to maximum derivative.

We now show some plots of the frequencies, $f^{(p n_r n_z)} = \omega^{(p n_r n_z)}/2\pi$ inferred from the simulations, for various modes $(p n_r n_z)$ versus $k_z$ at fixed $k_{p,n}$ with these latter values obtained by the above procedures. Also shown are the frequencies predicted by the H - K expression, Eq. (1.3), for the same wave-vector components and a demagnetization coefficient of $N_\parallel = 0.098$.

The $\Delta$ symbols in Fig. 5.2 show the results of the OOMMF simulation for the $(00 n_z)$ modes, as a function of $k_z$ in units of $\pi/h$, for which the lowest frequency is 7.81GHz. Not included are the accompanying cap-mode frequencies which are both 6.41GHz within the resolution. The continuous curve shows the frequencies predicted by the Herring-Kittel (H - K) formula. The agreement is surprisingly good, especially at small $k_z$ where H - K is expected to break down. To some extent this may arise from the fact that the internal magnetic field is position dependent, being lower at the caps, and thereby producing an effect partially compensating that of $N_\perp$. The $\square$ symbols show the predictions of the one-dimensional Schrodinger equation as described in section 2.2.

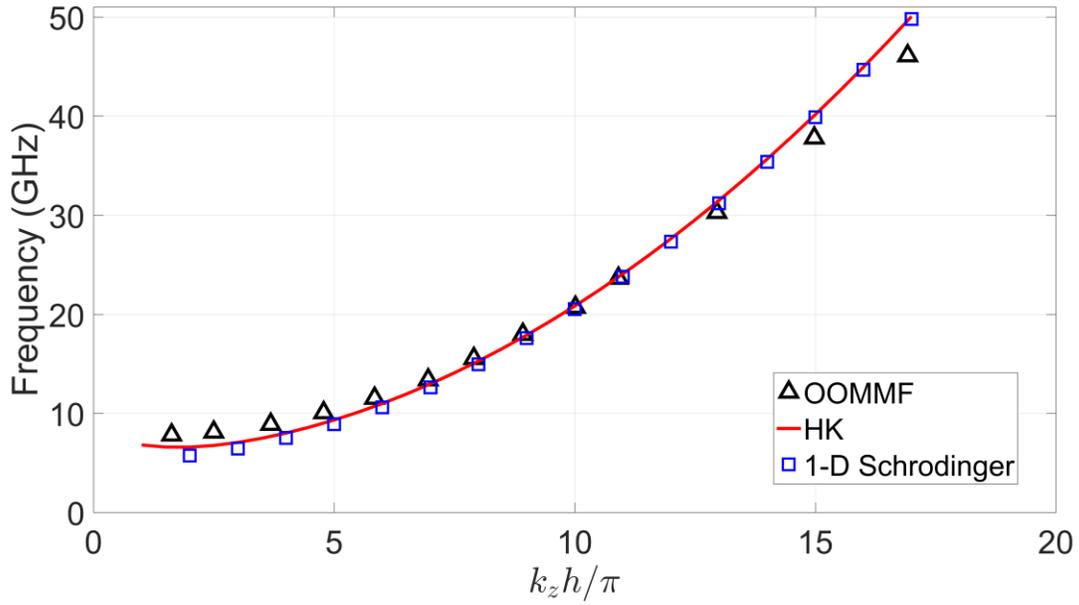

**Fig. 5.2. The frequencies of the $(0\,0\,n_z)$ modes vs. $k_z$ as obtained from OOMMF simulations ($\Delta$ symbols), Herring-Kittel formula (continuous line) and the Schrodinger equation ($\square$), as discussed in section 2.2. For the Schrodinger equation data $k_z h/\pi$ is a proxy for $n_z + 2$.**

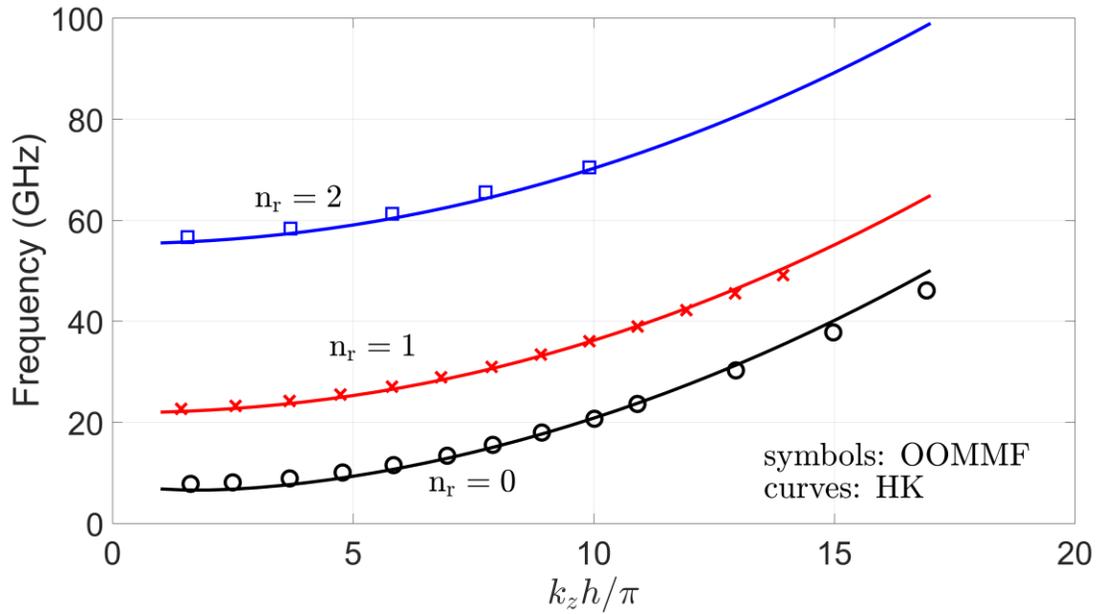

**Fig. 5.3. The symbols show the OOMMF simulations for the mode frequencies versus the dimensionless wave-vector $k_z$ for $p = 0$ and $n_r = 0,\ 1$ and $2$; the continuous curves show the predictions of the H – K expression.**

The symbols in Fig. 5.3 show the OOMMF simulations of $p = 0$ mode frequencies as a function of the dimensionless $k_z$ for $n_r = 0$, $1$ and $2$ while the continuous curves show the predictions for the corresponding mode numbers of the H - K formula. Again, the agreement is excellent. Readers may notice that some modes are missing in these plots. This is because they were not excited with the protocols used, but we are confident they exist and that their mode patterns conform with the general framework presented here.

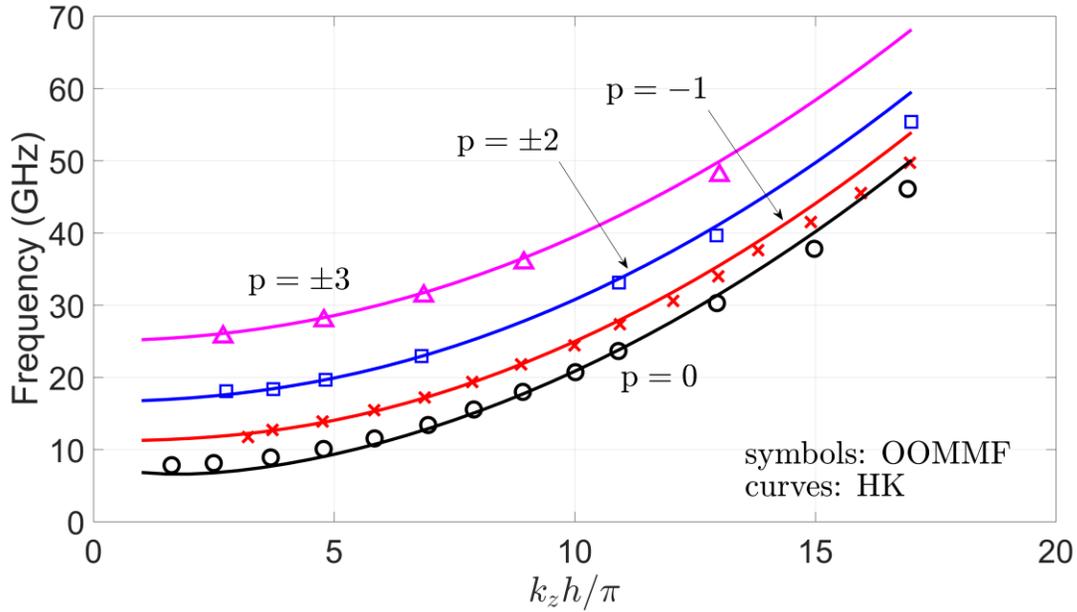

**Fig. 5.4. The OOMMF mode frequencies vs. $k_z$ for $n_r = 0$ and $p = 0,\ -1,\ \pm 2,\ \pm 3$. The curves show the predictions of the H – K equation.**

Lastly, Fig. 5.4 shows the OOMMF simulations for mode frequencies vs. $k_z$ for $n_r = 0$ and $p = 0,\ -1,\ \pm 2,\ \pm 3 \cdots$. The H - K formula again gives an excellent overall representation.

Analogous to our approximating the position dependence of the bulk states with a form $\cos(k_z z)$ we can use $e^{-\kappa z}$ and $e^{\kappa(z-h)}$ to qualitatively describe the amplitude in the vicinity of the cylinder faces for the cap mode; i.e., we take k as imaginary by writing $k = i\kappa$, in which case $k^2$ is replaced by $-\kappa^2$ in a Herring-Kittel like expression, which pushes the frequency *below* that of the first extended mode.

### 5.1. Why the H - K formula works so well

The H - K relation is sometimes referred to as a spin-wave dispersion relation, and indeed it is tantamount to saying that the normal modes of the body can be described by a continuous (or quasicontinuous) variable k. As explained in Appendix A, this assertion is grossly incorrect when the spatial variation of **m** is on a scale comparable to the dimensions of the body[56]; the mode functions must then take account of the *shape* of the body, and be described by discrete sets of appropriate mode numbers, which may or may not be wavevector-like.

Nevertheless, if it should turn out that some discrete modes with slow spatial variation are well described by wavevector-like variables, then for those modes the H - K expression can be expected to give the frequency to good approximation, especially if one is in the exchange dominated limit. This is the situation in the present investigation. We have seen that for every bulk mode, we can identify a reasonably well-defined $k_z$ and an orbital angular momentum quantum number p. We also see a radial dependence in **m**(r) that pretty well matches a Bessel function $J_p(k_\perp r)$ from which can obtain a $k_\perp$.

That the mode functions should look this way is not an accident. We have some support for this functional behavior from the theory of the infinite cylinder in the exchange dominated limit. We intend to publish the details of this theory separately, and here we only summarize the key results. In first order perturbation in the small parameter

$$\zeta = \frac{M_0}{D_{ex}k^2} \tag{5.1}$$

we find that the mode function is given by

$$\begin{pmatrix} m_x \\ m_y \end{pmatrix} = \frac{1}{\sqrt{2}} \left[ \left( J_p(k_\perp r) + f^-(r) \right) e^{i(p\varphi - \omega t)} \begin{pmatrix} 1 \\ i \end{pmatrix} + f^+(r) e^{i((p+2)\varphi - \omega t)} \begin{pmatrix} 1 \\ -i \end{pmatrix} \right] e^{\pm i k_z z} . \tag{5.2}$$

The corrections $f^\pm$ are both $O(\zeta)$. We add here that we discovered the properties of the mode functions by experimenting with OOMMF first, and developed the theoretical framework later.

## 6. Examples of mode patterns

Table II in section 4 lists approximately 90 modes which we have identified. We will now present some accompanying mode patterns which display various behaviors. All figures in this section pertain to YIG cylinders with a diameter d = 2a = 75 nm and a height h = 300 nm in a magnetic field $H_0$ = 2000 Oe.

There is a wealth of information in these figures as we now explain. They all depict various aspects of the frequency space Fourier amplitude $\mathbf{m}(\mathbf{r}, \omega)$, which is a complex vector, i.e., its x, y, and z components are all complex numbers. The real parts make a vector, and so do the imaginary parts, which we can call the real and imaginary parts of the complex vector, $\Re\mathbf{m}(\mathbf{r},\omega)$ and $\Im\mathbf{m}(\mathbf{r},\omega)$. We discard the z component leaving the xy projection $\mathbf{m}_\perp(\mathbf{r}, \omega)$. Circular panels such as Figs. 6.1a and 6.1b show xy cross sections of the cylinder, while rectangular panels such as Fig. 6.1c show yz cross sections through a diameter of the cylinder. In the circular panels, the arrows show the directions and relative magnitudes of either $\Re\mathbf{m}_\perp(\mathbf{r},\omega)$ or $\Im\mathbf{m}_\perp(\mathbf{r},\omega)$, while the thin black lines show contour levels of the magnitudes of these same vectors, i.e., either $\left|\Re\mathbf{m}_\perp(\mathbf{r},\omega)\right|$ or $\left|\Im\mathbf{m}_\perp(\mathbf{r},\omega)\right|$. These contour levels are also color coded according to the scale on the right. As explained in Sec. 3.3 — see Eqs. (3.4) and (3.5) — the $\Re$ panel shows the spins a quarter-cycle after the $\Im$ panel. That is, time proceeds from $\Im$ to $\Re$, which is reflected in the sequence of panels a) and b) when both $\Re$ and $\Im$ parts are shown. In the rectangular panels we show contours of $\Re m_x(\mathbf{r},\omega)$; again, the contours are color coded. Because the circular panels show the magnitude of the vector in the xy plane, while the rectangular panels show only the x component, the contour levels in the two types of panels cannot be directly compared. Furthermore, depending on just how a particular mode is excited, the spins can have larger projections along the x or y directions at the particular time captured in the xy cross sections, and this can further affect the values of the contour levels in the rectangular panels vis-a-vis the circular ones. The most salient feature is the variation or the relative Fourier amplitude within a panel. We add here that our simulations are done with 50 vertical layers of cells of height 6 nm each. There is a layer extending from z = 144 to 150 nm, and another from z = 150 to 156 nm. Hence, circular panels such as in Figs. 6.1a and 6.1b that are labeled z = 147 nm correspond to the midpoint of the cell layer just below the midplane of the cylinder; panels labeled z = 3 nm such as in Fig. 6.2a show the lowest layer; panels labeled z = 39 nm such as in Figs. 6.5a and 6.5b show the 7th layer from the bottom. However, in the text and figure captions we have described the panels at z = 3 nm and z = 147 nm as lying at z = 0 and z = h/2, as this is more natural and intuitively easier to understand.

In the interest of clarity, we shall repeat these points as necessary, and add further information about the patterns as we discuss them one by one.

We start with the lowest lying modes: the nominal bulk uniform precession or cylindrical Kittel (0, 0, 0) mode with f = 7.813 GHz, which is concentrated within the body of the cylinder, away from the caps, together with the even (0, 0, g) and odd (0, 0, u) cap modes concentrated on the top and bottom cylinder faces with a mean frequency of 6.543 GHz and a splitting that is too small to be resolved.

Fig. 6.1a shows $\Im\mathbf{m}_\perp^{(000)}(x, y, z = h/2, \omega)$ while Fig. 6.1b shows $\Re\mathbf{m}_\perp^{(000)}(x, y, z = h/2, \omega)$ for the (0, 0, 0) bulk mode with f = 7.813 GHz; here $\Im\mathbf{m}_\perp(x, y, z = h/2, \omega)$ and $\Re\mathbf{m}_\perp(x, y, z = h/2, \omega)$ denote the normalized vector fields of the Fourier amplitude given by following prescriptions:

$$\frac{\Im\mathbf{m}_\perp(\mathbf{r_i}, \omega)}{\left[ N_{cell}^{-1} \sum_i |\Im\mathbf{m}(\mathbf{r_i}, \omega)|^2 \right]^{1/2}} \times K \qquad \text{and} \qquad \frac{\Re\mathbf{m}_\perp(\mathbf{r_i}, \omega)}{\left[ N_{cell}^{-1} \sum_i |\Re\mathbf{m}(\mathbf{r_i}, \omega)|^2 \right]^{1/2}} \times K \qquad (6.1)$$

where K is a global constant scale factor whose value is chosen as 1000 for convenience for all mode patterns (this and subsequent ones). The arrows indicate the xy projection of the magnetization. The lines and color coding depict the contours of constant amplitude, either $\Re\mathbf{m}_\perp(\mathbf{r_i}, \omega)$ or $\Im\mathbf{m}_\perp(\mathbf{r_i}, \omega)$. Ideally these would be concentric circles but there is always contamination at some level from other modes. There may also be numerical errors associated with the discretization. Time proceeds from $\Im$ (panel (a)) to $\Re$ (panel (b)), a quarter cycle later.

Fig. 6.1c shows $\Re m_x^{(000)}(x = 0, y, z, \omega)$, again with contour lines together with color coding. Note the contour lines are quite parallel to the faces, a behavior that arises from the strong influence of exchange in these small samples and validates the factorized form Eq. (2.21) for $F_-(r, z)$.

Fig. 6.2a shows $\Re\mathbf{m}_\perp^{(00g)}(x, y, z = 0, \omega)$ for the symmetric (g) cap mode with f = 6.543 GHz. Figs. 6.2b and 6.2c show $\Re m_x^{(00g)}(x = 0, y, z, \omega)$ and $\Re m_x^{(00u)}(x = 0, y, z, \omega)$ for the even (g) and odd (u) cap modes respectively. We see that the mode intensity is strongly concentrated

near the cylinder faces, dropping off rapidly as one proceeds to the interior. Note the anti-symmetric character of the u mode is clearly apparent as seen from the node at $z = h/2$.

Figs. 6.3a, b show $\Im\mathbf{m}_\perp^{(-10u)}(x,y,z=0,\omega)$ and $\Re\mathbf{m}_\perp^{(-10u)}(x,y,z=0,\omega)$ of the $(-1, 0, u)$ antisymmetric 10.06 GHz cap mode which has a node at $r = 0$. Note how the spins wind through an angle of $2\pi$ as we proceed counter-clockwise around the line $r = 0$. Fig. 6.3c shows $\Re m_x^{(-10u)}(x=0,y,z,\omega)$ where the antisymmetric behavior in z is evident.

Figs. 6.4a, b show $\Im\mathbf{m}_\perp^{(10u)}(x,y,z=0,\omega)$ and $\Re\mathbf{m}_\perp^{(10u)}(x,y,z=0,\omega)$ for the neighboring $p = +1$ mode with a frequency of 10.35 GHz. Here one encounters the "retrograde" motion associated with the oppositely winding sense of m with the azimuthal angle $\varphi$.

We next consider a mode with multiple nodes along z. As remarked earlier the splitting between $\pm p$ modes diminishes as the overall mode numbers increase, so we designate them with both signs since our mode projection method generally yields a superposition. Here we consider the $(\pm 1, 0, 5)$ mode(s) with $f = 17.29$ GHz. Figs. 6.5a, b show $\Im\mathbf{m}_\perp^{(\pm 105)}(x,y,z=39\text{nm},\omega)$ and $\Re\mathbf{m}_\perp^{(\pm 105)}(x,y,z=39\text{nm},\omega)$ while Fig. 6.5c shows $\Re\mathbf{m}_\perp^{(\pm 105)}(x=0,y,z,\omega)$. (We recall that $z = 39$ nm is the midplane of the 7th layer of cells from the bottom of the cylinder.) Note the mode patterns in the xy plane now display nodes since the modes with $p = +1$ and $p = -1$ interfere to form a partial standing wave. Our plot in the yz plane contains the two end nodes arising from orthogonality to the cap modes discussed above (the patterns for which we do not show) as well as the five interior nodes. If we use the projection technique described in Appendix B, we can again separate the two modes. This is shown in Figs. 6.6a and 6.6b where we plot $\Re\mathbf{m}_\perp^{(+105)}(x,y,z=39\text{nm},\omega)$ and $\Re\mathbf{m}_\perp^{(-105)}(x,y,z=39\text{nm},\omega)$. These are also the modes for which we resolved the splitting via the beat pattern in Fig. 3.2a.

As an example of a mode with a larger azimuthal mode number and mixed p character, Fig. 6.7 shows a plot of $\Re\mathbf{m}_\perp^{(\pm 305)}(x,y,z=39\text{nm},\omega)$, which has a frequency of 31.25 GHz.

Finally, we present a mode with additional radial nodes. Such a mode will have a high frequency considering the relatively small diameter of our sample. Figs. 6.8a and 6.8b show $\Re\mathbf{m}_{\perp}^{(020)}(x, y, z = h/2, \omega)$ and $\Re m_x^{(020)}(x = 0, y, z, \omega)$ for the (020) mode with a frequency of f = 56.64 GHz.

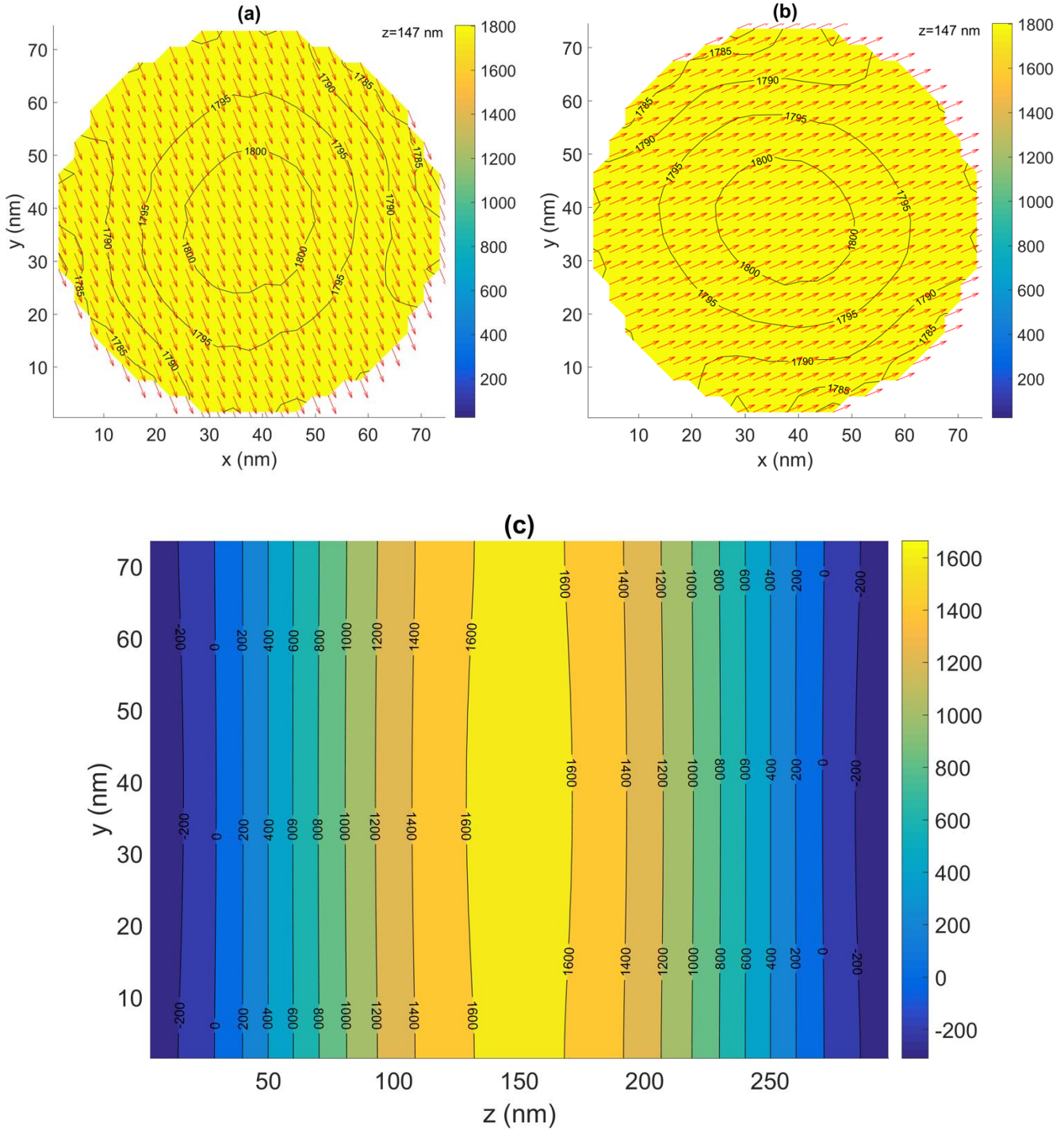

**Fig. 6.1.** Mode pattern for the lowest (Kittel-like) bulk (0 0 0) mode, with a frequency of 7.813 GHz. Parts a) and b) show an x - y cross section of the imaginary and real parts of the Fourier transform amplitude through the cylinder mid point while part c) is the real y - z cross section containing the cylinder axis. The arrows show the direction of the spins. The spin orientations for the real part correspond to a time 1/4 cycle later than that for the imaginary part. The lines show contours of constant $|\mathbf{m}(\mathbf{r}, \omega)|$ in (a) and (b) and of constant $\mathbf{m}_X(\mathbf{r}, \omega)$ in (c); these values are also color coded according to the scale given to the right.

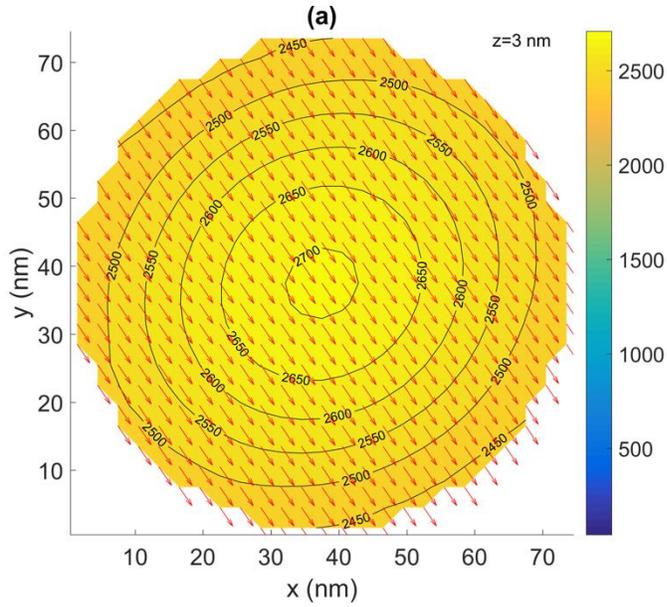

**Fig. 6.2.** The p = 0, n$_r$ = 0 cap modes. Part a) shows the real x - y cross section at z = 0 of the (00g) symmetric cap mode; this mode has the globally lowest frequency of 6.543 GHz. Parts b) and c) show the real y - z cross sections of the g and u modes containing the cylinder axis of the from which the surface confinement is apparent.

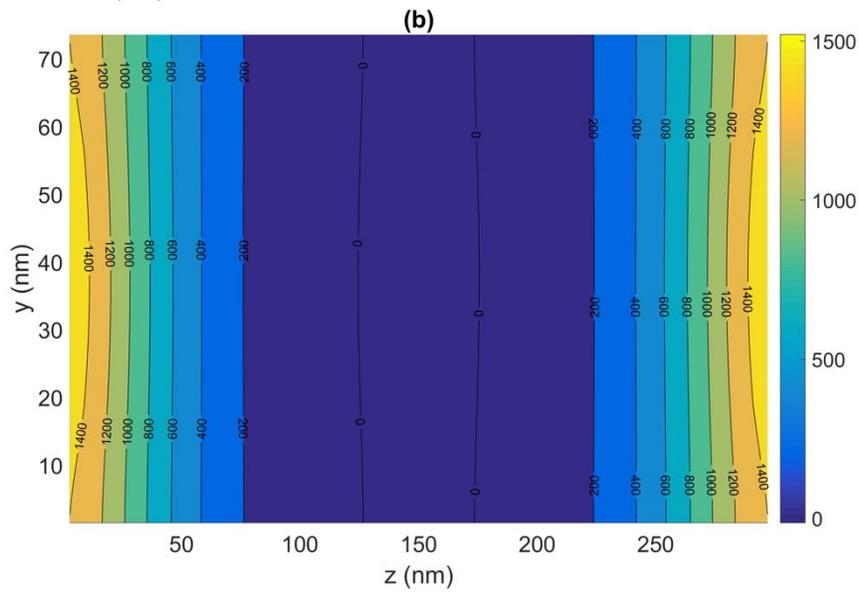

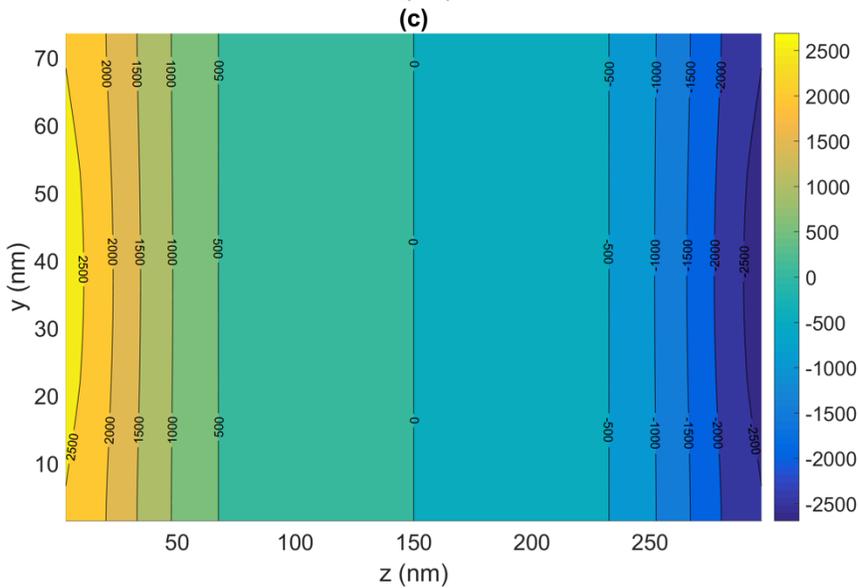

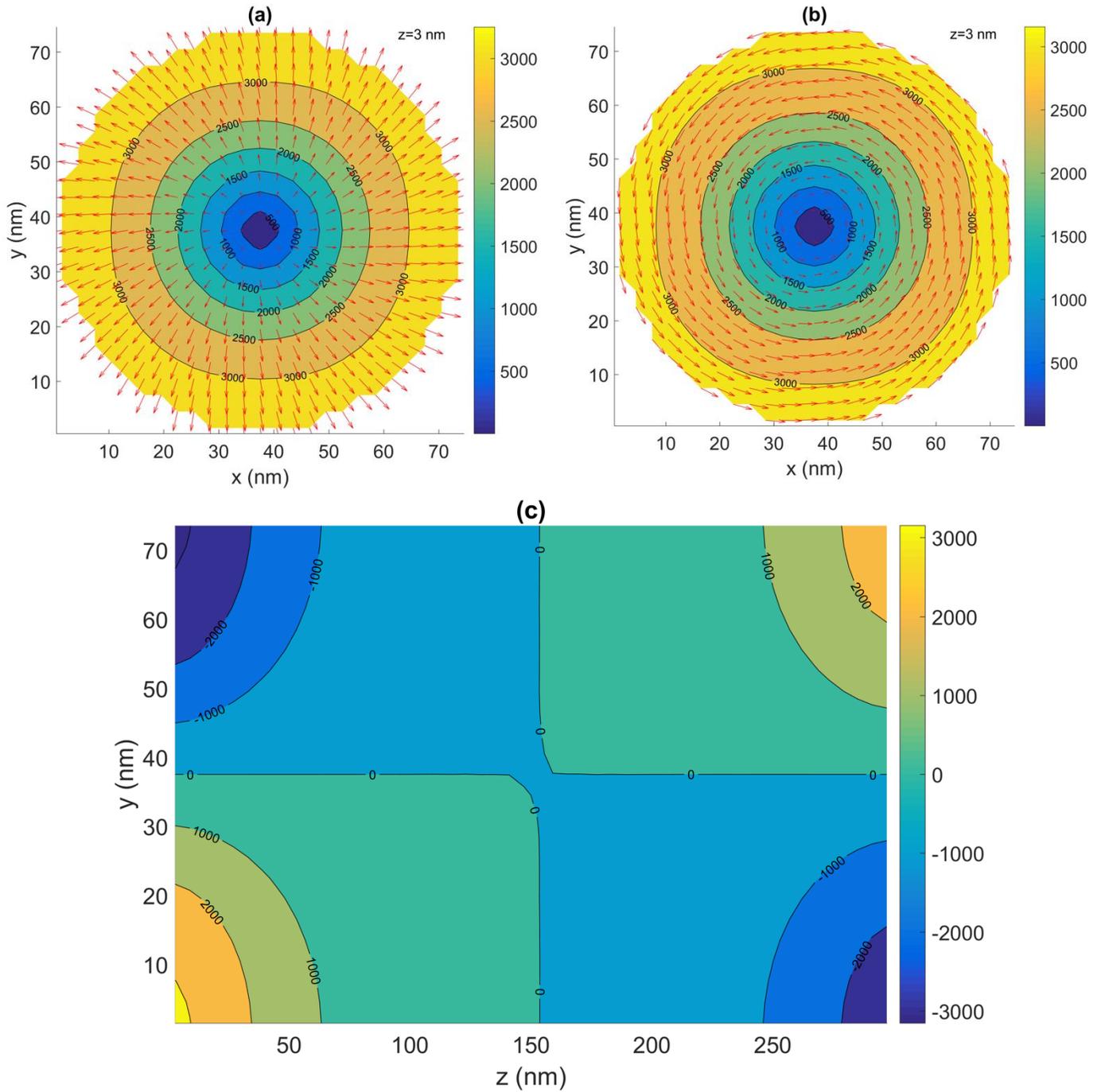

**Fig. 6.3. Parts a) and b) show the imaginary and real x - y cross sections at z = 0 of the (−1 0 u) antisymmetric cap mode with a frequency of 10.06 GHz. Part c) is the real y - z cross section containing the cylinder axis from which the surface confinement is apparent.**

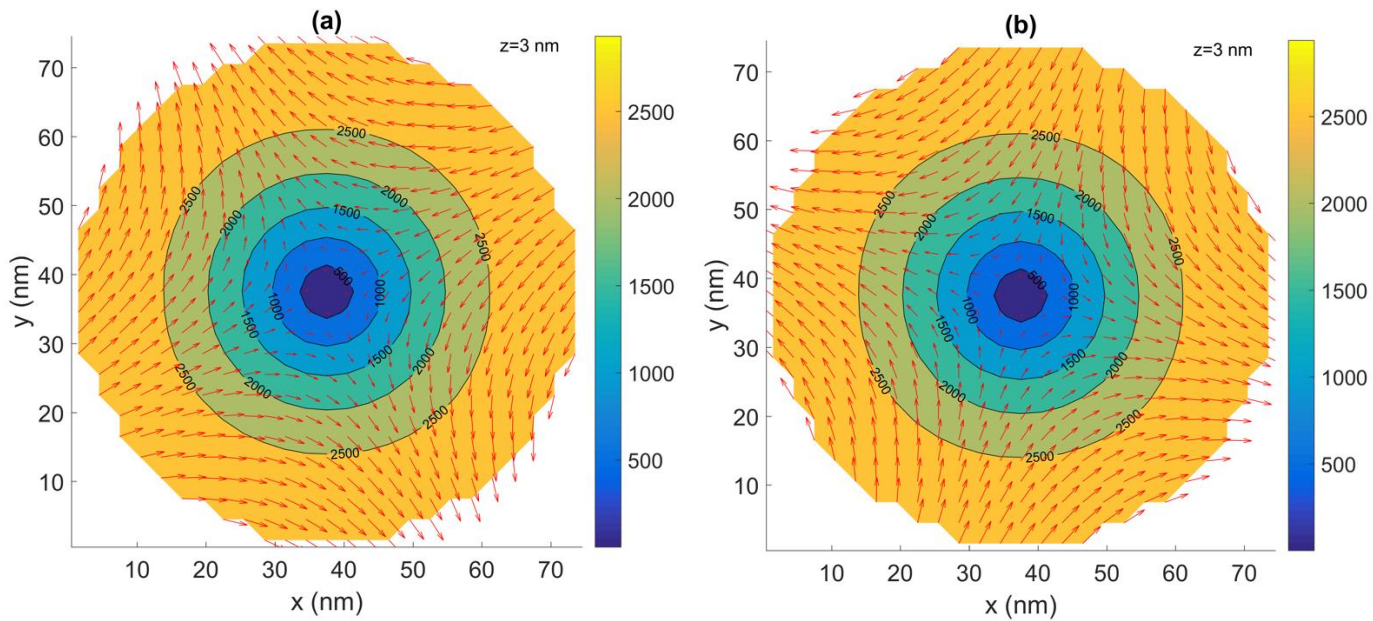

**Fig. 6.4. Parts a) and b) show the imaginary and real x - y cross sections at z = 0 of the (10u) antisymmetric cap mode with a frequency of 10.35 GHz. Important to note here is the retrograde character of the spin winding.**

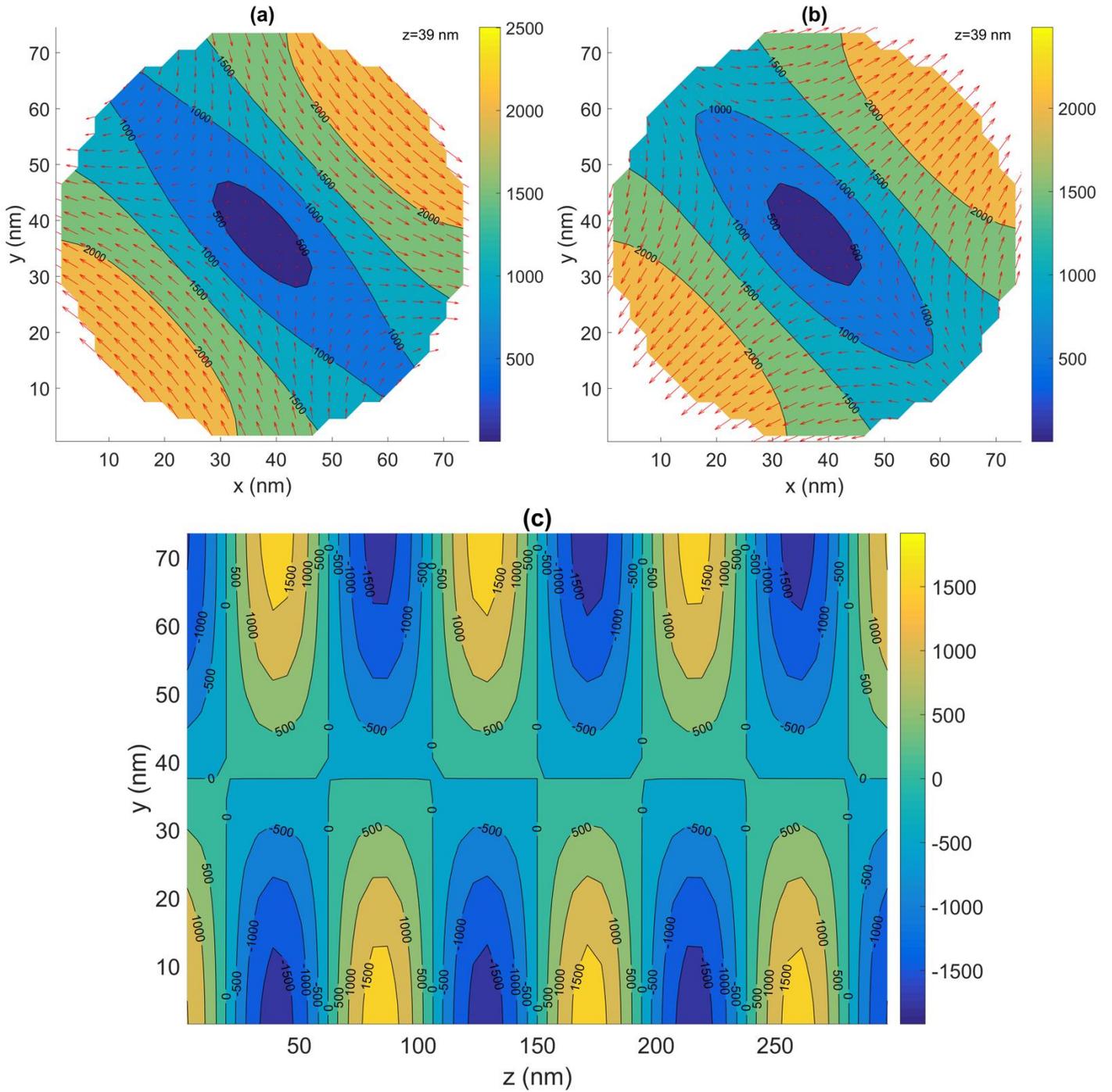

**Fig. 6.5. Parts a) and b) show the imaginary and real x - y cross sections at z = 39 nm of the (±105) modes with a frequency of 17.29 GHz. Note the standing wave behavior of these cross-sections arising from the superposition of azimuthally counter propagating modes. However, at a fixed point in space the spins still precess counterclockwise. Part c) is the real y - z cross section containing the cylinder axis.**

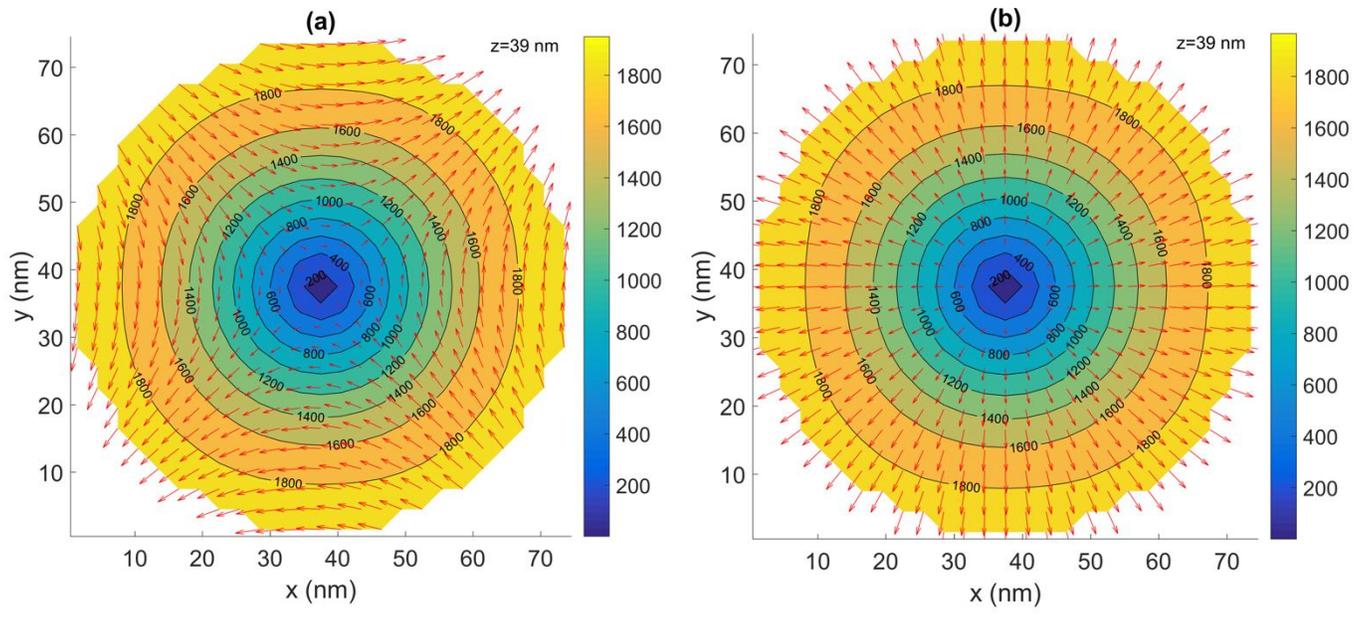

**Fig. 6.6.** Here the projection technique described in Appendix B has been used to separate the modes with p = +1 and p = −1 from the same data used to construct figure 6.5. Note the azimuthal intensity is now constant as appropriate for running waves.

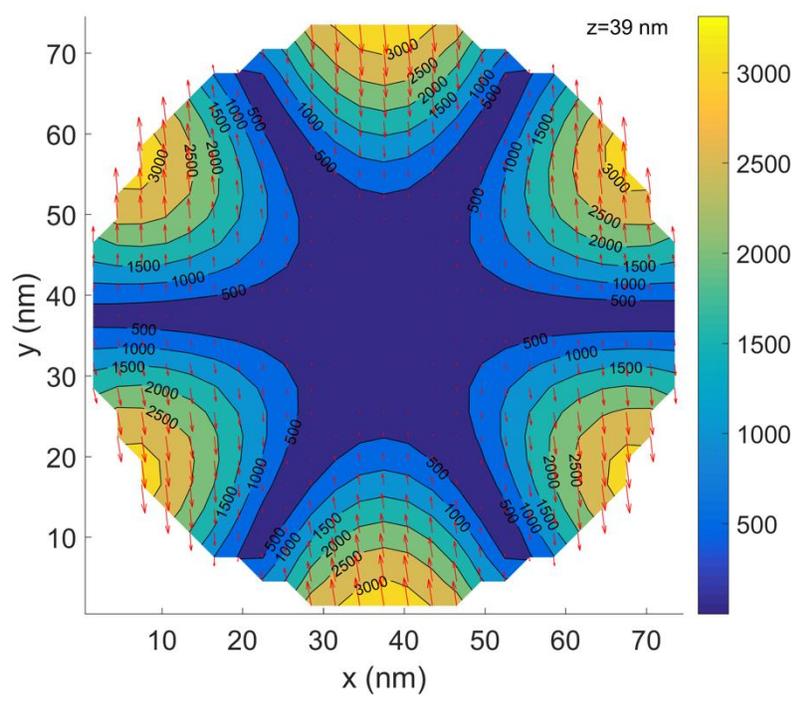

**Fig. 6.7.** The real x - y cross section of the (±305) bulk mode with a frequency of 31.25 GHz showing a multiplicity of azimuthal nodes. Note also the deep central node due to the $r^3$ behavior of $J_3(k_\perp r)$.

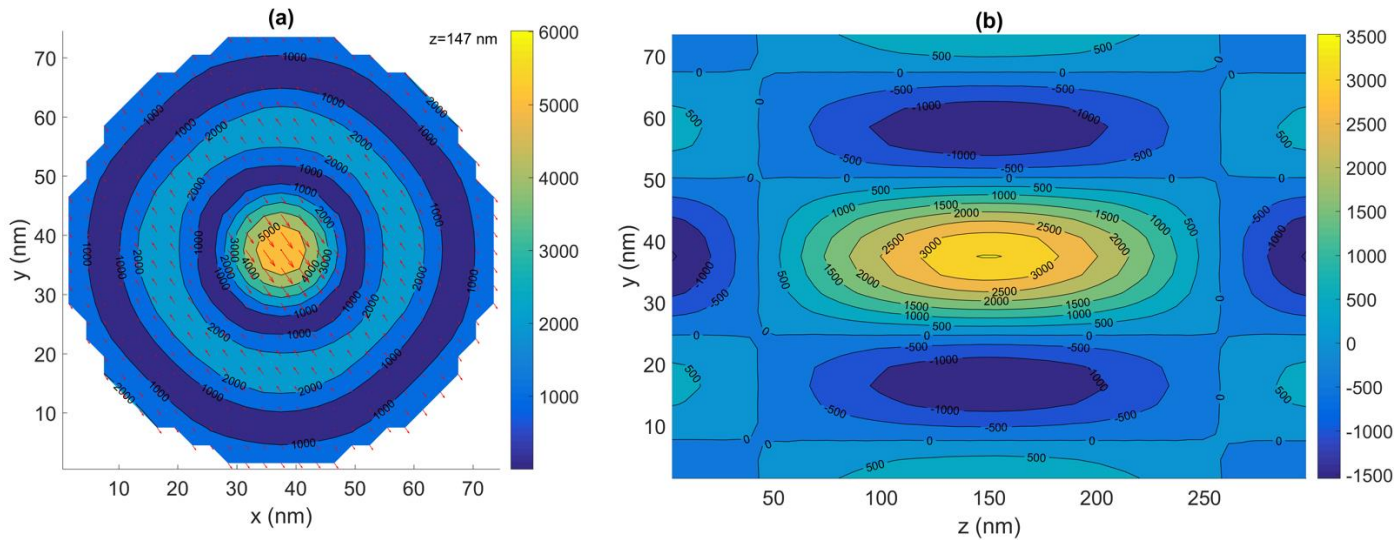

**Fig. 6.8. The real x-y and y-z cross-sections of the (020) mode with a frequency of 56.64 GHz exhibiting multiple radial nodes.**

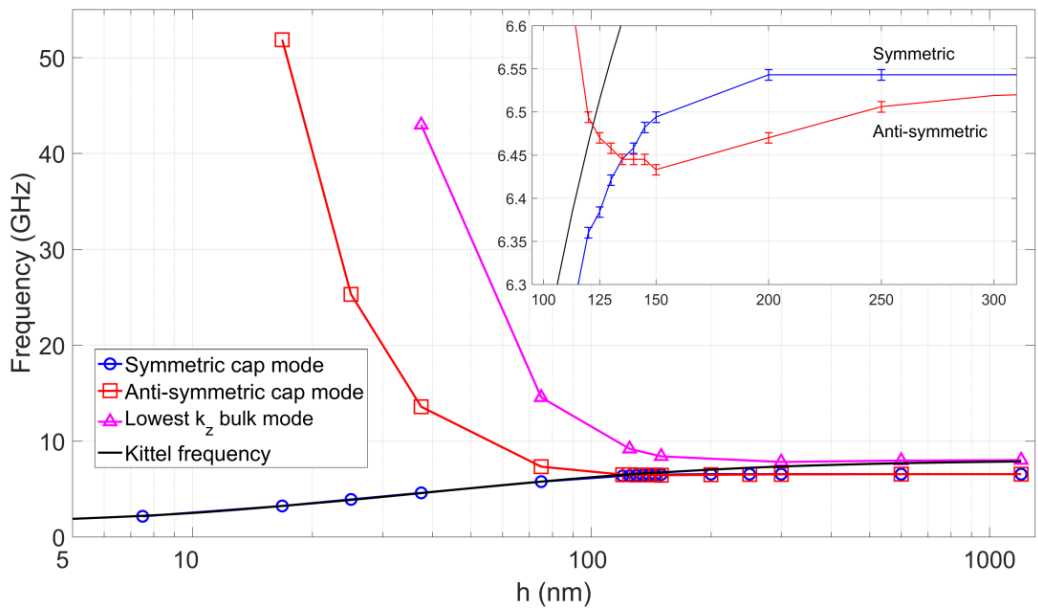

**Fig. 7.1. The dependence of the frequency on h for the lowest $k_z$ bulk mode (000), symmetric cap mode (00g), and antisymmetric cap mode (00u) of a YIG cylinder with d =75nm. Also shown is the frequency predicted by the Kittel expression. The inset shows the region where the symmetric cap mode crosses over the antisymmetric cap mode for small h. In the region below about 100nm the (00g) mode replaces the (000) mode as the quasi uniform mode we associate with the Kittel formula. See discussion.**

# 7. Evolution of the low-lying mode behavior with cylinder height

Although the majority of our simulations for mode patterns were carried out for a YIG cylinder with d = 2a = 75 nm and h = 300 nm, the behavior of the three lowest lying p = 0 modes, (000), (00g), and (00u), was studied over the much wider range of h extending from 7.5nm to 1200 nm[57]. Fig. 7.1 shows the frequencies emerging from the OOMMF simulations together with the predictions based on the Kittel expression, Eq. (1.2), according to the demagnetization coefficients found by Joseph and Schlomann[58]. Importantly we see that the (00g) "cap mode" *evolves into the dominant mode in the thin-disc limit*, while the (000) bulk mode becomes dominant (although lying slightly higher in frequency) in the long cylinder limit. The Kittel formula, which is a single equation, actually describes two different modes in these limits and does not apply well to any mode for 150 nm $\lesssim$ h $\lesssim$ 300 nm. The level crossing of the (00g) and (00u) modes does not violate the Wigner-von Neumann anti-crossing theorem due to their even and odd character. Furthermore, the apparent violation of the ordering of energy levels of 1d Schrodinger equations is resolved by noting that the dipole-dipole interaction is a *non-local* perturbation which puts the eigenvalue problem outside the Sturm-Liouville class.

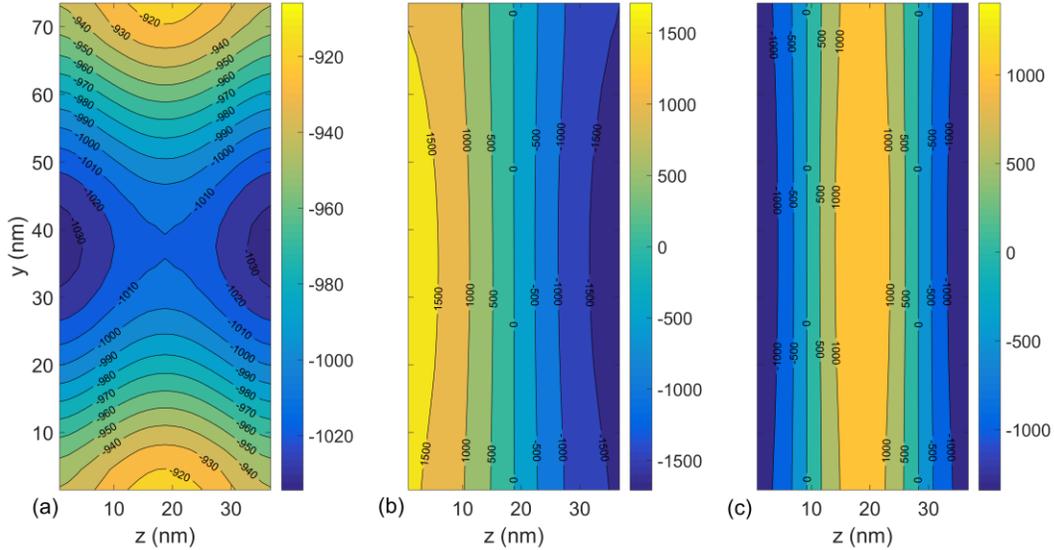

**Fig. 7.2. The contours of constant intensity for three modes of a YIG cylinder with d = 75 nm and h = 37.5 nm. a) The 4.59 GHz symmetric (00g) cap mode, corresponding to the Kittel mode in the thin-disc limit; note it has no nodes. b) The 13.57 GHz antisymmetric (00u) cap mode. c) The 42.97 GHz (000) "bulk" mode.**

It is clear from this discussion that the true behavior of the dominant FMR response cannot be described by a simple Kittel-like expression; a characterization in terms of demagnetization coefficients is often inadequate and glosses over the spatial complexity of the mode with the greatest spectral weight in a uniformly excited sample.

Some mode patterns for a YIG cylinder with d = 75 nm and h = 37.5 nm are shown in figure 7.2. The cell size for these simulations is 3 x 3 x 1.5 nm^3. Fig7.2a shows the shows the 4.59 GHz symmetric (00g) "cap" mode which for these dimensions has become the dominant mode. It is a cap mode in name only since it spans the entire sample. Rather than the planar constant amplitude contours encountered in the longer cylinders, this mode now has approximately cylindrical ones. Figure 7.2b shows the antisymmetric cap mode for a cylinder with the same dimensions which now has a significantly higher frequency of 13.57 GHz; here the contours of constant amplitude run approximately parallel to the faces indicating that the odd mode has substantially changed character from the even one. Lastly, we show the bulk mode in Fig. 7.2c. Consistent with the results in Fig. 7.1 this mode has the highest frequency, 42.97 GHz. The contours of constant amplitude run approximately parallel to the faces and we obtain a quite regular sine wave with one complete wavelength along z.

# 8. Conclusions and possible applications

We have explored the mode structure of nanoscopic cylinders of yttrium iron garnet, both analytically and through many-spin simulations, in a regime where the effects of exchange dominate the response. In addition to the extended (bulk) modes, which can be classified in terms of the azimuthal, radial, and axial mode numbers designated as $p, n_r, n_z$, we find symmetric and antisymmetric combinations of cap modes, that are localized at each of the cylinder faces. In all cases they lie lower in frequency than the accompanying family of modes $n_z$ for given azimuthal and radial mode numbers $p$ and $n_r$. When examining the height dependence, we find that the dominant FMR response *cannot* be precisely described by a simple Kittel-like formula. How this picture would change in passing to the dipole dominated limit is deserving of additional study.

By way of applications, there is a growing effort directed at using magnetic bits for computation as well as data storage. In particular it has been demonstrated that the required logic operations can be accomplished with lines and arrays of dipole-coupled, single-domain, bar magnets with dimensions of a few hundred nanometers[59]. While promising, this approach is still restricted to what can be done using binary macro-spin flips (and cascades thereof) of the individual magnets.

Looking farther ahead it is natural to ask if logic functions can be performed by exploiting the *internal* dynamics of nano particles. For the case of wave-guide based operations this is already a world-wide activity[60]. But here we envision exciting (and mixing) large amplitude resonant modes within a single nanoparticle. This can involve a single or multiple inputs applied simultaneously or sequentially, with different microwave frequencies and/or polarizations. When the particles are small the various modes are well separated and can be addressed individually and rapidly. By exploiting intrinsic nonlinearities and optimizing the sample dimensions (to tune the mode frequencies), different pump frequencies can be efficiently mixed, a topic we are exploring independently. Progress in this direction requires an understanding of the low amplitude mode structure of the particles involved.

Apart from the cap modes, the modes studied here are all extended in character and in that sense are the standing wave counter part of the plane wave modes that were the starting point of the non-linear analysis carried out by Suhl discussed in the introduction. However ongoing simulations of nano-scale cylinders and elliptical discs at large precession amplitudes display instabilities involving the edge-nucleation of dynamic vortex and antivortex modes. Possibly related instabilities occur at domain walls in stripes[61,62]. The connection between such states and low-lying extended modes in understanding large amplitude precession dynamics in nanomagnets is currently unclear.

## Acknowledgement

This research was carried out under the support of U.S. Department of Energy through grant DE-SC0014424. This research was supported in part through the computational resources and staff contributions provided for the Quest high performance computing facility at Northwestern University which is jointly supported by the Office of the Provost, the Office for Research, and Northwestern University Information Technology.

## Appendix A: The Herring - Kittel equation

The Herring - Kittel (H - K) expression, given earlier as Eq. (1.3), is

$$\omega^2 = \gamma^2(H_{in,z} + D_{ex}k^2)(H_{in,z} + 4\pi M_0 \sin^2\theta + D_{ex}k^2) \tag{A.1}$$

where we wrote $H_{in,z} \equiv H_0 - 4\pi N_\parallel M_0$ with $N_\parallel$ an axial demagnetization coefficient, and

$$D_{ex} = \frac{2A_{ex}}{M_0}, \tag{A.2}$$

which fixes the exchange energy density used in the OOMMF simulation,

$$E_{ex} = \frac{A_{ex}}{M_0^2} \sum_{i,j} \frac{\partial M_j}{\partial x_i} \frac{\partial M_j}{\partial x_i}. \tag{A.3}$$

To understand the remarkable agreement between our OOMMF results and this formula, let us recall how it is derived.

The linearized Landau-Lifshitz equation is

$$\frac{d\mathbf{m}}{dt} = \gamma \hat{\mathbf{z}} \times \left( H_{in,z}\mathbf{m} - M_0\mathbf{h} + \frac{2A_{ex}}{M_0}\nabla^2\mathbf{m} \right) \tag{A.4}$$

where $\mathbf{h}$ is the dynamic demagnetization (or dipolar) field induced by $\mathbf{m}$. Fourier transforming with respect to space and assuming a time dependence, $e^{-i\omega t}$ we obtain

$$-i\omega\mathbf{m}_k = \gamma \hat{\mathbf{z}} \times \left( H_{in,z}\mathbf{m}_k - M_0\mathbf{h}_k - \frac{2A_{ex}}{M_0}k^2\mathbf{m}_k \right). \tag{A.5}$$

The field $\mathbf{h}$ is governed by the Maxwell equations,

$$\nabla \cdot \mathbf{h} = -4\pi\nabla \cdot \mathbf{m}, \quad \nabla \times \mathbf{h} = 0, \tag{A.6}$$

together with the requirements that the normal component of $\mathbf{b} = \mathbf{h} + 4\pi\mathbf{m}$ and the tangential component of $\mathbf{h}$ be continuous at the boundary of the particle. If we now Fourier transform Eq.'s (A.6), and simply *ignore* the effects of the boundary conditions we obtain

$$\mathbf{h}_k = -4\pi \frac{\mathbf{k} \cdot \mathbf{m}_k}{k^2} \mathbf{k} . \tag{A.7}$$

Inserting (A.7) into Eq. (A.5), requiring the equations for the resulting two components to be compatible, and taking $H_{in,z}$ to be homogeneous, leads immediately to the HK formula, Eq. (A.1).

From the above derivation we see that, qualitatively, the HK approximation accounts for the axial (static) demagnetization but neglects some part of the transverse contribution. Since the dipole-dipole interaction is long ranged, this neglect is qualitatively profound, and quantitatively valid only for short wavelengths, when $ka \gtrsim 1$. When this condition is satisfied we can argue that the magnetic charges induced on the "lateral" surface by $\mathbf{m}$ change sign rapidly on a length scale $k^{-1}$, so the field produced by them dies off on the same length scale and may be ignored in the bulk of the particle. To further clarify this behavior, we will derive the HK formula in a second way.

As is known from magnetostatics, the (normalized) magnetic field, $\mathbf{h}(\mathbf{r})$, can be regarded as arising from a magnetic charge density $\nabla \cdot \mathbf{m}(\mathbf{r})$ according to

$$\mathbf{h}(\mathbf{r}) = -4\pi \int_{V^+} \frac{(\mathbf{r} - \mathbf{r}')}{|\mathbf{r} - \mathbf{r}'|^3} \nabla' \cdot \mathbf{m}(\mathbf{r}') d^3 r'. \tag{A.8}$$

The integral here is taken to extend infinitesimally beyond the particle volume, as indicated by the superscript in $V^+$. In this way both volume and surface charges are included. If we Fourier transform this equation, we obtain

$$\mathbf{h_k} = -4\pi \int d^3 k' \frac{\mathbf{k}' \cdot \mathbf{m_{k'}}}{k'^2} \mathbf{k}' \, \chi_{\mathbf{k} - \mathbf{k}'} \tag{A.9}$$

where $\chi_\mathbf{k}$ is the Fourier transform of unity over the particle volume:

$$\chi_\mathbf{k} = \frac{1}{(2\pi)^3} \int_{V^+} e^{-i\mathbf{k} \cdot \mathbf{r}} d^3 r \ . \tag{A.10}$$

We can write (A.9) more compactly as

$$\mathbf{h}_k = -4\pi \left( \frac{\mathbf{k} \cdot \mathbf{m_k}}{k^2} \mathbf{k} \right) * \chi_\mathbf{k} \tag{A.11}$$

where $*$ denotes a convolution. For a cylinder of height h and radius a,

$$\chi_\mathbf{k} = \frac{\pi a^2 h}{(2\pi)^3} \frac{\sin k_z h}{k_z h} \frac{2 J_1(k_\perp a)}{k_\perp a} \ . \tag{A.12}$$

When $k_z h \gtrsim 1$ and $k_\perp a \gtrsim \pi$ in the convolution in Eq. (A.11), $\chi_\mathbf{k}$ can be approximated by a delta-function, $\delta(\mathbf{k})$, and we recover Eq. (A.7). For smaller k this approximation is invalid.

## Appendix B.  Relation between standing and running waves in $\varphi$

The magnetization fields corresponding to running waves associated with orbital angular momentum p and $-p$ (with $p > 0$) have the form,

$$\mathbf{m_p}\left(\mathbf{r},t\right) = F\left(r,z\right)\left[\cos\left(p\varphi - \omega t\right)\hat{\mathbf{x}} - \sin\left(p\varphi - \omega t\right)\hat{\mathbf{y}}\right], \qquad (B.1a)$$

$$\mathbf{m_{-p}}\left(\mathbf{r},t\right) = F\left(r,z\right)\left[\cos\left(-p\varphi - \omega t\right)\hat{\mathbf{x}} - \sin\left(-p\varphi - \omega t\right)\hat{\mathbf{y}}\right], \qquad (B.1b)$$

where $F(r,z)$ is an unspecified function. By superposing these fields, we obtain a standing wave pattern,

$$\mathbf{m_p^s}\left(\mathbf{r},t\right) = \mathbf{m_p} + \mathbf{m_{-p}} = F\left(r,z\right)\cos\left(p\varphi\right)\left[\cos\left(\omega t\right)\hat{\mathbf{x}} + \sin\left(\omega t\right)\hat{\mathbf{y}}\right]. \qquad (B.2)$$

We wish to recover the running wave patterns from a knowledge of the standing wave pattern. To do this, we first transform the latter by rotating the amplitude by $\pi/2p$, and the vector by direction by $\pi/2$. The transformed field is,

$$\begin{aligned}\mathbf{m_p^T}\left(\mathbf{r},t\right) &= \mathbf{m_p^s}\left(r,z,\varphi - \left(\pi/2p\right), t + \left(\pi/2\omega\right)\right) \\ &= F\left(r,z\right)\sin\left(p\varphi\right)\left[-\sin\left(\omega t\right)\hat{\mathbf{x}} + \cos\left(\omega t\right)\hat{\mathbf{y}}\right]. \end{aligned} \qquad (B.3)$$

It is now easy to see that the difference of the transformed and the original stationary wave pattern gives us $\mathbf{m_p}$:

$$\mathbf{m_p}\left(\mathbf{r},t\right) = \left[\mathbf{m_p^s}\left(\mathbf{r},t\right) - \mathbf{m_p^T}\left(\mathbf{r},t\right)\right]\Big/2 . \qquad (B.4a)$$

Likewise, the sum gives $\mathbf{m_{-p}}$:

$$\mathbf{m_{-p}}\left(\mathbf{r},t\right) = \left[\mathbf{m_p^s}\left(\mathbf{r},t\right) + \mathbf{m_p^T}\left(\mathbf{r},t\right)\right]\Big/2 . \qquad (B.4b)$$

What this means is the following. Supposing OOMMF has produced a pattern which has p nodal lines in $\varphi$ at some fixed time. We denote this pattern by $\mathbf{m_p^s}$ as above. We then consider the vector fields

$$\begin{pmatrix} m_{\pm p,x}\left(r,z,\varphi\right) \\ m_{\pm p,y}\left(r,z,\varphi\right) \end{pmatrix} = \frac{1}{2}\begin{pmatrix} m_{p,x}^s\left(r,z,\varphi\right) \\ m_{p,y}^s\left(r,z,\varphi\right) \end{pmatrix} \mp \frac{1}{2}\begin{pmatrix} -m_{p,y}^s\left(r,z,\varphi - \pi/2p\right) \\ m_{p,x}^s\left(r,z,\varphi - \pi/2p\right) \end{pmatrix} . \qquad (B.5)$$

If these combinations are used as initial conditions in OOMMF, they should evolve into $+p$ and $-p$ running waves as indicated. In this way we can obtain positive confirmation that these running waves are indeed eigenmodes; this is how we separated the $+p$ and $-p$ modes shown in Fig.'s 6.5 and 6.6.

# References

[1] J. H. E Griffiths, Anomalous High-frequency Resistance of Ferromagnetic Metals, Nature, **158**, 670 (1946), DOI: 10.1038/158670a0.

[2] H. Suhl, Subsidary Absorption Peaks in Ferromagnetic Resonance at High Signal Levels, Phys. Rev. **101**, 1437 (1956).

[3] H. Suhl, The Theory of Ferromagnetic Resonance as High Signal Powers, J. Phys. Chem. Of Solids **1**, 209-227 (1957).

[4] M. J. Donahue and D. G. Porter, OOMMF User's Guide, Version 1.0, **NISTIR 6376**, National Institute of Standards and Technology, Gaithersburg, MD (Sept 1999).

[5] R. D. McMichael and M. D. Stiles, Magnetic Normal Modes of Nanoelements, J. Appld. Phys. **97**, 10590 (2005); DOI: 10.1063/1.1852191.

[6] Jian-Gang Zhu, Xiaochun Zhu, and Yuhui Tang, Microwave assisted magnetic recording, IEEE Trans. Magn. **44**, 125-131 (2007).

[7] K. Rivkin, M. Benakli, N. Tabat, and H. Yin, Physical principles of microwave assisted magnetic recording, J. Appl. Phys. **115**, 214312 (2014).

[8] S. Okamoto, N. Kikuchi, M. Furuta, O. Kitakami and T. Shimatsu, Microwave assisted magnetic recording technologies and related physics J. Phys. D: Appl. Phys. **48**, 3530011 (2015).

[9] K. Rivkin, N. Tabat, and S. Foss-Schroeder, Time-dependent fields and anisotropy dominated magnetic media, Appl. Phys. Lett. **92**, 153104 (2008).

[10] Jinho Lim, Zhaohui Zhang, Anupam Garg, and John Ketterson, Simulating Resonant Magnetization Reversals in Nanomagnets, IEEE Trans. Magn. **57**, 1300304 (2021) DOI: 10.1109/TMAG.2020.3039468.

[11] Jinho Lim, Zhaohui Zhang, Anupam Garg and John B. Ketterson, Pi Pulses in a Ferromagnet: Simulations for Yttrium Iron Garnet, J. Mag. Mag. Mater. **527**, 167787 (2021). DOI: https://doi.org/10.1016/j.jmmm.2021.167787

[12] Eiichi Fukushima and Stephan B. W. Roeder, Experimental Pulsed NMR: A Nuts and Bolts Approach, CRC Press, 1981.

[13] Christophe Thirion, Wolfgang Wernsdorfer, and Dominique Mailly, Switching of magnetization by nonlinear resonance studied in single nanoparticles, Nature Materials, Vol **2**, Pg. 524 (2003); DOI:10.1038/nmat946.

[14] H. Sato, T. Kano, T. Nagasawa, K. Mizushima, and R. Sato, Magnetization switching of a Co/Pt multilayered perpendicular nanomagnet assisted by a microwave field with time varying frequency, Phys. Rev. Applied **9**, 054011 (2018).

[15] Y. Li, et. al., Nutation spectroscopy of a nanomagnet driven into deeply nonlinear ferromagnetic resonance, Phys. Rev. X **9**, 041036 (2019).

[16] For a recent example of zero-field resonances in a nano-scale object see: Wonbae Bang, F. Montoncello, M. T. Kaffash, A. Hoffmann, J. B. Ketterson, and M. B. Jungfleisch, Ferromagnetic resonance spectra of permalloy nano-ellipses as building blocks for complex magnonic lattices, J. Appl. Phys. **126**, 203902 (2019); https://doi.org/10.1063/1.5126679.

[17] If technical issues associated with growing high quality YIG films on non-magnetic substrates can be solved (i.e., not GGG), achieving long mode lifetimes at low temperatures is anticipated. Nanocylinders can then be patterned and excited by superconducting antennas wherein close coupling can be achieved together with locally tailored field profiles to more effectively couple to specific modes.

[18] A. Yamaguchi, K. Motoi, A. Hirohata, H. Miyajima, Y. Miyashita, and Y. Sanada, Broadband ferromagnetic resonance of $Ni_{81}Fe_{19}$ wires using a rectifying effect, Phys. Rev. B **78**, 104401 (2008), DOI: 10.1103/PhysRevB.78.104401.

[19] U. Ebels, J. L. Duvail, P. E. Wigen, L. Piraux, L. D. Buda, and K. Ounadjela, Ferromagnetic resonance studies of Ni nanowire arrays, Phys. Rev. B **64**, 144421 (2001), DOI: 10.1103/PhysRevB.64.144421.

[20] C. A. Ramos, M. Vazquez, K. Nielsch, K. Pirota, J. Rivas, R. B. Wehrspohn, M. Tovar, R. D. Sanchez and U. Gösele, J. Mag. Mag. Mater. **272–276**, Part 3, 1652 (2004), doi.org/10.1016/j.jmmm.2003.12.233.

[21] A del Campo and C Greiner, SU-8: A photoresist for high-aspect-ratio and 3D submicron lithography, J. Micromech. Microeng. **17**, R81 (2007) doi:10.1088/0960-1317/17/6/R01.

[22] L. D. Landau and E. M. Lifshitz, Course in Theoretical Physics, Statistical Physics Part II Pergamon Press, Oxford, 1980, Section 69.

[23] C. Kittel, On the Theory of Ferromagnetic Resonance Absorption, Phys. Rev. **73** (2): 155–161 (1948), doi:10.1103/PhysRev.73.155.

[24] Exciting such modes with a uniform microwave field requires they have an even number of nodes, otherwise the net interaction averages to zero. To address this problem, highly efficient antennas having multiple elements that are spaced so as to spatially resonate with some set of modes have recently been employed: Jinho Lim, Wonbae Bang, Jonathan Trossman, Andreas Kreise, Matthias Benjamin Jungfleisch, Axel Hoffmann, C. C. Tsai, and John B. Ketterson, Direct detection of multiple backward volume modes in yttrium iron garnet at micron scale wavelengths, Phys. Rev. B **99**, 014435 (2019).

[25] F. Bloch, On the theory of ferromagnetism, Z. Physik **61**, 206 (1930); On the Theory of the Exchange Problem and the Appearence of Retentive Ferromagnetic, Z. Physik **74**, 295 (1932).

[26] A. M. Clogston, H. Suhl, L. R. Walker and P. W. Anderson, Possible Source of Line Width in Ferromagnetic Resonance, Phys. Rev. **101**, 903 (1956).

[27] A. M. Clogston, H. Suhl, L. R. Walker and P. W. Anderson, Ferromagnetic Line Width in Insulating Materials, J. Phys. Chem. Solids **1**, 129 (1956).

[28] J. E. Mercereau and R. P. Feynman, Physical Conditions for Ferromagnetic Resonance, Phys. Rev. **104**, 63 (1956).

[29] L. R. Walker, Magnetostatic Modes in Ferromagnetic Resonance, Phys. Rev. **105**, 390 (1957).

[30] R. W. Damon and J. R. Eshbach, Magnetostatic modes of a ferromagnet slab, J. Phys. Chem. Solids **19**, 308 (1961), doi.org/10.1016/0022-3697(61)90041-5.

[31] R. W. Damon and H. Van De Vaart, Propagation of Magnetostatic Spin Waves at Microwave Frequencies in a Normally-Magnetized Disk, J. Appl. Phys. **36**, 3453 (1965); https://doi.org/10.1063/1.1703018

[32] D. E. De Wames and T. Wolfram, Dipole-exchange spin waves in ferromagnetic films, J. Appl. Phys. **41**, 987 (1970); doi.org/10.1063/1.1659049.

[33] R. E. Arias, Spin-wave modes of ferromagnetic films, Phys. Rev. B **94**, 134408 (2016).

[34] R. I. Joseph, and E. Schlomann, Theory of magnetostatic modes in long, axially magnetized cylinders, J. Appl. Phys. **32**, 1001 (1961).

[35] R. Arias and D. L. Mills, Theory of spin excitations and the microwave response of cylindrical ferromagnetic nanowires, Phys. Rev. B **63**, 134439 (2001).

[36] C. Herring and C. Kittel, On the theory of spin waves in ferromagnetic media, Phys. Rev. **81**, 869 (1951).

[37] S. Shultz and G. Dunifer, Observation of spin waves in sodium and potassium, Phys. Rev. Lett. **18**, 283 (1967).

[38] N. Masuhara, D. Candela, D. O. Edwards, R. F. Hoyt, H. N. Scholz, D. S. Sherrill and R. Combescot, Collisionless Spin Waves in Liquid $^3$He, Phys. Rev. Lett. **53** (12), 1168 (1984).

[39] V. P. Silin, Oscillations of a Fermi-liquid in a magnetic field, Zh. Eksp. Teor. Fiz. **33**, 1227 (1957) [Sov. Phys. JETP **6**, 945 (1958)].

[40] A. J. Leggett, Spin diffusion and spin echoes in liquid $^3$He at low temperature, J. Phys. C: Solid State Phys. **3**, 448 (1970).

[41] G. Venkat, D.Kumar, M. Franchin, O. Dmytriiev, M. Mruczkiewicz, H. Fangohr, A.Barman, M. Krawczyk, and A. Prabhakar, Proposal for a Standard Micromagnetic Problem: Spin Wave Dispersion in a Magnonic Waveguide IEEE Trans. Magn. **49**, 524–529 (2013).

[42] L. D. Landau and E. M. Lifshitz, Statistical Physics Part II, Pergamon Press, Section 69; the form written here follows Herring and Kittel.

[43] This argument is standard, and can be found in many books. See, e.g., R. Shankar, Principles of Quantum Mechanics, Plenum, New York, 1980, Exercise 12.5.1 and Fig. 12.1.

[44] See, e.g., Anupam Garg, Classical Electromagnetism in a Nutshell, Princeton University Press, Princeton, N. J., 2012, pp. 174–177, and 239–240.

[45] See Eqs. (19), (23), and (24) in Ref. 34.

[46] See the unnumbered equation two equations above Eq. (22) in Ref. 29.

[47] H. A. Bethe and E. E. Salpeter, Quantum Mechanics of One and Two Electron Atoms, Springer, Berlin (1957), Sections 29 and 30.

[48] W. S. Ament and G. T. Rado, Electromagnetic Effects of Spin Wave Resonance in Ferromagnetic Metals, Phys. Rev. **97**, 1558 (1955)

[49] Amikam Aharoni, Introduction to the Theory of Ferromagnetism, Oxford, 1996, p. 178.

[50] See, e.g., C. M. Bender and S. Orszag, Advanced Mathematical Methods for Scientists and Engineers, McGraw-Hill, 1978, Chapter 9.

[51] This has the virtue of replacing the discrete OOMMF field profile with a continuous one; it also gave somewhat better agreement with the simulated OOMMF mode profiles.

[52] M. Sato and Y. Ishii, Simple and approximate expressions of demagnetizing factors of uniformly magnetized rectangular rod and cylinder, J. Appl. Phys. **66** (2), 983 (1989). DOI: 10.1063/1.343481.

[53] A. G. Gurevich and G. A. Melkov, Magnetization oscillations and waves, CRC Press, Boca Raton, 1996.

[54] It can also drive additional modes having some overlap on the chosen mode at some initial instant in time.

[55] Alternatively, we could Fourier transform the simulated form of $\overline{m_r(z,r,\omega)}$ along z, where the bar indicates a radial average, and identify the peak in the spectrum. Since our description is only semi quantitative this procedure was not attempted.

[56] In particular Eq. (A.11) cannot be replaced by Eq. (A.7) in Appendix A. In this regard, we find the discussion by L. R. Walker, Resonant Modes of Ferromagnetic Spheroids, J. Appl. Phys. **29**, 318-324 (1958) and his Fig. 6 to be particularly germane.

[57] The simulations were done with cell dimensions of 3 nm × 3nm in the x and y directions, and the dimension $\Delta z$ along the z direction was adjusted so as to give varying numbers of layers depending on the height h. The number of layers varied from 10 for h = 7.5 nm, to 25 for h= 16.8 to 145 nm, and either 25 or 50 for h ≥ 150nm. The choice of 25 layers is enough to

capture the spatial variation of the three modes we are seeking in this section. It is more important to increase the run time (by as much as a factor of 8) in order to resolve the modes in frequency.

[58] R. I. Joseph and E. Schlomann, Demagnetizing field in non-ellipsoidal bodies, J. Appl. Phys. **36**, 1579 (1965).

[59] For a review see: M. T. Niemier, G. H. Bernstein, G. Csaba, A. Dingler, X. S. Hu, S. Kurtz, S. Liu, J. Nahas, W. Porod, M. Siddiq and E. Varga, Nanomagnet logic: progress toward system-level integration, J. Phys. Condens. Matter **23,** 493202 (2011).

[60] Abdulqader Mahmoud, Florin Ciubotaru, Frederic Vanderveken, Andrii V. Chumak, Said Hamdioui, Christoph Adelmann, and Sorin Cotofana, Introduction to spin wave computing, J. Appl. Phys. **128**, 161101 (2020); https://doi.org/10.1063/5.0019328

[61] D. J. Clarke, O. A. Tretiakov, G.-W. Chern, Ya. B. Bazaliy, and O. Tchernyshyov, Dynamics of a vortex domain wall in a magnetic nanostrip: Application of the collective-coordinate approach, Phys. Rev. B **78**, 134412 (2008).

[62] M. Klaui, Topical Review: Head-to-head domain walls in magnetic nanostructures, J. Phys. Condens. Matter **20,** 313001 (2008).